# A Solid-State Nanopore Signal Generator for Training Machine Learning Models

*Jaise Johnson[1], Chinmayi R Galigekere[2] and Manoj M Varma[1*]*

[1]Center for Nanoscience and Engineering, Indian Institute of Science, Bangalore, Karnataka, India – 560012.

[2]Indian Institute of Science Education and Research, Thiruvananthapuram, Kerala, India – 695551.

[*]Email: mvarma@iisc.ac.in

**Abstract:** Translocation event detection from raw nanopore current signals is a fundamental step in nanopore signal analysis. Traditional data analysis methods rely on user-defined parameters to extract event information, making the interpretation of experimental results sensitive to parameter choice. While Machine Learning (ML) has seen widespread adoption across various scientific fields, its potential remains underexplored in solid-state nanopore research.

In this work, we introduce a nanopore signal generator capable of producing extensive synthetic datasets for machine learning applications and benchmarking nanopore signal analysis platforms. Using this generator, we train deep learning models to detect translocation events directly from raw signals, achieving over 99% true event detection with minimal false positives.
.

## 1. Introduction

Nanopore devices, widely used for DNA and protein sequencing, consist of two reservoirs of ionic solution connected by a nanometer-scale pore in a thin membrane. When an electric field is applied, analyte molecules translocate through the pore, causing characteristic changes in the ionic current. The first demonstration of single-molecule characterization using biological nanopores i.e. protein-based pores was pioneered by Kasianowicz et al. in 1996 [1]. This was followed by Akeson et al. in 2000 [2], who showed that nanopores could distinguish between pyrimidine and purine segments in RNA. In parallel, solid-state nanopores (SSNPs), typically fabricated on silicon or silicon nitride membranes [3,4], emerged as an alternative due to their greater stability and tuneable pore dimensions.

SSNP-based measurements rely on analyzing ionic current signals to detect and classify translocation events, typically characterized by current blockade patterns. Many software tools, such as Translyzer [6], EasyNanopore (ESY) [7], AutoNanopore (AN) [8], and EventPro (EP) [9], have been developed to process nanopore data. However, effective usage of these tools requires domain expertise, as parameter selection significantly affects data interpretation. Additionally, the sheer volume of nanopore data necessitates faster processing pipelines, leading to efforts in parallelizing data analysis on CPUs [10]. As nanopore sensing expands into commercial applications, non-parametric event detection methods are needed to improve automation and accessibility. Moreover, as this field is progressing towards sequencing with complex events where information is encoded in the event shapes [11], classical methods of nanopore signal processing becomes challenging.

Machine learning (ML) [12] has transformed various fields, from weather prediction and autonomous vehicles to medical imaging. Sequencing using biological nanopore systems, already employs ML in base calling [13]. ML-based approaches offer a promising non-parametric alternative for SSNP signal processing while also benefiting from parallelization on GPUs, potentially bringing massive improvement in processing speeds. However, training robust ML models require large, diverse datasets representative of real-world nanopore signals. The scarcity of publicly available datasets, combined with the high cost at the current state of the technology and time-consuming nature of nanopore experiments, has hindered the development of ML-driven nanopore analysis tools.

In this work, we present a nanopore signal generator capable of generating large, controlled datasets to capture the variability in nanopore signals. We use these generated synthetic datasets to train a deep learning model for event detection on both experimental and additional generated synthetic data. Furthermore, we benchmark the performance of a few widely used nanopore signal analysis software tools using our generated dataset to showcase the utility of the generated datasets for benchmarking different software.

## 2. Signal Generator

The components of an SSNP signal can be broadly classified into three, the baseline, translocation events and noise. The baseline in the signal denotes the current level resulting from open pore current. Translocation events are the dips in current value due to the partial blockage of the pore when an analyte passes through the pore. Noise in SSNP signals is of different types 1/f noise, dielectric noise, white noise, amplifier noise [14,15], and AC powerline noise. In addition to the noise, the events are also deformed and broadened by the finite filter response of the nanopore systems as well as the mismatch in timescales of the translocation events with the sampling frequencies. Accounting to these nuances in nanopore data we developed a signal generator with two versions for different applications:

1. For generating datasets containing signal segments of specified lengths for Machine Learning.
2. For generating long nanopore signals that mimic an experimental signal. This version is accompanied by a Graphical User Interface (GUI) for better accessibility. Supporting Figure SF2-SF4 shows different pages in the GUI.

Other than how the signals are saved, both versions follow a similar method to generate signals. The user can tune different parameters to adjust the amplitude and pulse widths of the events, event density of the signal, generate multilevel events [16] while also choosing the distribution and density of events and adding slow drifts to the signals. We have also accounted for the finite response of the nanopore system due to capacitive effects (RC effects) and also a low pass filter to account for the timescale mismatch as mentioned earlier.

A summary of the options available in the generator is described below as a workflow.

1. Specify the minimum and maximum values for pulse widths and amplitudes of the desired events. These parameters are passed to random number generators that follow a uniform distribution, resulting in two arrays: one for pulse widths and another for amplitudes. If multi-level events are required, users can either explicitly specify the amplitude and width of each level, or simply define the number of levels, in which case the corresponding amplitudes and widths are randomly generated. Based on this information, events are generated in the next step.
2. Specify the desired distribution for event locations, options include logistic, uniform, or exponential (with exponential as the default). Additionally, define the desired event density within the signal using the event density factor in the program. This factor determines the percentage of data points in the entire signal that correspond to translocation events. The events generated in the previous step are then added to a zero baseline according to the specified distribution and density.
3. If desired, users can introduce slowly varying baseline drifts into the signal. This is achieved by concatenating multiple sinusoidal and square wave components. The program allows users to specify the amplitude and number of harmonics for the sinusoidal waves, as well as the number and duration of steps in the square wave pulses.
4. The next step involves modeling pulse deformation caused by the filter rise time [17], which results from RC effects in the system. To simulate this, the entire signal is passed through a numerical model of an RC low-pass filter. The resistance (R) and capacitance (C) values can be specified by the user to control the extent of signal smoothing.
5. If the option to do a low pass filtering is chosen the RC filtered signal is passed through a low pass filter.
6. Once this signal is generated, we add noise White noise using a uniform random number distribution using *random* library in python, Color noises using *colorednoise* package in python which is based on an algorithm provided by J Timmer and M Koenig [18]. The noise level is adjusted by specifying the *nstd* parameter in the program.

A graphical illustration of the process of signal generation can be seen in Figure 1. For the first version the data is saved as a set of .csv files with each of the file name containing a unique name with an index in the specified path. For the second version of the program, once the signal is generated, the signal is saved as a .csv file along with a version of clean signal. These files can be loaded in Clampfit by axon instruments to convert it into *.abf* file to perform analysis with popular nanopore analysis software. Another file with event information such as event start point, end point, event width, amplitude, number of levels in an event etc., named - *Signalname-details.csv* is also saved to the specified path. This file can be used to cross-reference and evaluate the results of each of the software to compare and benchmark its performance. In addition, we developed a graphical user interface (GUI) for the simulator to make it accessible to users without programming experience. Screenshots of the GUI is provided in the supporting information.

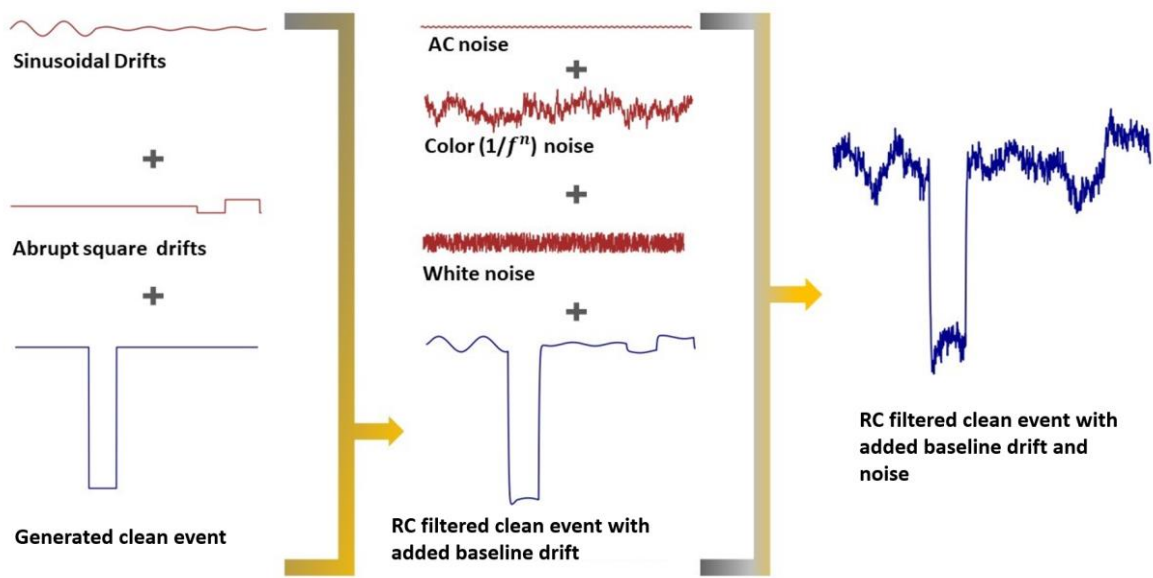

*Figure 1*: *Illustration of process of generation of the signals.*

## 3. Deep Learning Models for Translocation Event Identification

We designed a deep neural network [19,20] to identify event containing regions in nanopore signals. Specifically, we implemented a 1D convolutional neural network (CNN) [21] based on the ResNet architecture [22] to classify signal segments as containing events or not. A detailed description of the network architecture is provided in the Supporting Information.

The model processes signal segments of 1024 data points, outputting a scalar value indicating the presence (1) or absence (0) of events. To support robust feature learning, we generated a dataset containing signals with diverse noise levels, multiple event amplitudes, and varying event densities. The dataset comprised 40,900 signal segments, split into training (70%), validation (20%), and test (10%) subsets. Signals containing events were labelled 1, while those without were labelled 0. Illustration of a representative input batch is shown in Figure SF 5 of S.I.

We developed two deep learning models for event detection:
1. Baseline Model: Accepts raw nanopore signal segments as a single-channel input.
2. Enhanced Model: Utilizes a nine-channel input consisting of the raw signal along with eight additional pre-processed versions, enabling the model to leverage richer contextual and statistical information.

The preprocessing steps is described below.

3.1 Preprocessing with Block Averaging and Block Minimum

To enhance feature extraction, we applied block-based transformations to the raw signal. Specifically, we used:
- Block Averaging: The time series was divided into non-overlapping blocks of sizes 256, 128, 64, and 32, with each block replaced by its mean value.
- Block Minimum: Using the same block sizes, each block was also represented by its minimum value.

These transformations yielded eight derived signals four block-averaged and four block-minimum capturing both smoothed and extremal features across multiple temporal resolutions. Combined with the original signal, each input segment was represented as a nine-channel tensor. All segments were of fixed length 1024, resulting in input tensors of shape [64,9,1024], where 64 denotes the batch size. For comparison, the baseline model's input consisted of a single-channel tensor of shape [64,1,1024]. This preprocessing of signals is illustrated in Figure SF 7 of Supporting Information.

3.2 Model Training

The models were trained for 10 epochs using the Adam optimizer [23] on Google Colab GPUs. The learning rate was dynamically adjusted using the One-Cycle Learning Rate Policy [24]. To ensure effective learning and assess the model's stability, we repeated the training and testing using different random data splits, controlled by varying random seeds in PyTorch. Both the models achieved over 98% accuracy on both the test and validation datasets.

## 4. Datasets for Evaluation

To evaluate the performance of our models, we used two types of datasets: a synthetic dataset generated using our in-house signal simulator and a real-world experimental dataset.

4.1 Generated Synthetic Dataset for Performance Evaluation

While experimental data is commonly used to compare nanopore signal analysis tools, it limits the ability to measure absolute performance metrics, as the ground truth such as the exact number, amplitude, and width of events is typically unknown. In contrast, synthetic datasets offer fine-grained control over these parameters, enabling objective benchmarking.

We compiled a synthetic evaluation dataset of 105 signals generated using version 2 of our signal generator, covering a range of signal-to-noise ratios, baseline current levels, and event densities with each signal containing exactly 100 events. An illustration of the distribution of signals in terms of different generation parameters is shown in the supporting information. After excluding 9 signals that could not be opened by some software tools, the final dataset included 96 signals, which was used to compare the performances of the different programs.

4.2 Experimental Dataset for Performance Evaluation

To assess real-world performance and verify the applicability of models trained on synthetic data, we compiled an experimental dataset comprising three nanopore signals obtained from different sources. The signal *g10.abf* was provided by Sohini Pal. The other two signals, C*hipA.abf* and *ChipB.abf*, were obtained from the publicly available repository of the Drndić Lab and correspond to the experiments described in [25].

## 5 Evaluation on Synthetic and Experimental Datasets

We utilize the datasets mentioned above to compare the performances of the models to classical programs. The synthetic dataset enabled quantitative benchmarking against known ground truth, while the experimental dataset helped validate the practical applicability of our approach.

### 5.1 Testing on Generated Dataset

With known event locations and counts in the synthetic data, we compare the true and false detection rates of our models with classical software tools: Autonanopore, EasyNanopore, and EventPro in Section 5.1.1. Additionally, in section 5.1.2 we assess and compare the processing times of these programs.

In Section 5.1.3, we further examine the accuracy of classical tools in estimating event amplitude and width by reporting the average percent error. Our machine learning models were excluded from this comparison, as they are designed solely to detect the presence of events within fixed segments, without estimating precise event parameters.

The models process 1024-point signal segments. Hence the signals from the generated dataset were segmented accordingly. The true start and end points of events were binned into 1024-point windows, ensuring accurate detection assessment. This was necessary as non-overlapping segmentation could split events across two segments.

Since each synthetic signal contained 100 events, we optimized the parameters of the classical software to detect this known number of events per file, thereby minimizing false positives and ensuring a fair comparison. While the overall event counts were comparable across tools, some programs were more susceptible to misidentifying noise as events. To evaluate detection accuracy, we defined a ±10-point tolerance window around the true start and end positions of each event; any detections falling outside this window were considered false.

### 5.1.1 Percent of True/False Detection

The deep learning models demonstrated superior performance. The nine-channel model detected 98.89%, while the single-channel model detected 99.48%, averaging 99.2% across the generated data. The performance of the models trained on different distributions of training dataset generated by the randomly with the corresponding datasets are given in the table ST2 in the supporting information. EasyNanopore detected 80.48% of events, followed by EventPro (79.37%) and AutoNanopore (61.75%). Regarding false detections, the nine-channel model had the lowest rate at 2.04%, while the single-channel model had 3.516% averaging to 2.8%.

In contrast, classical methods had significantly higher false detection rates: EasyNanopore (19.51%), EventPro (20.62%), and AutoNanopore (38.25%). These results are illustrated in Figure 2a and 2b.

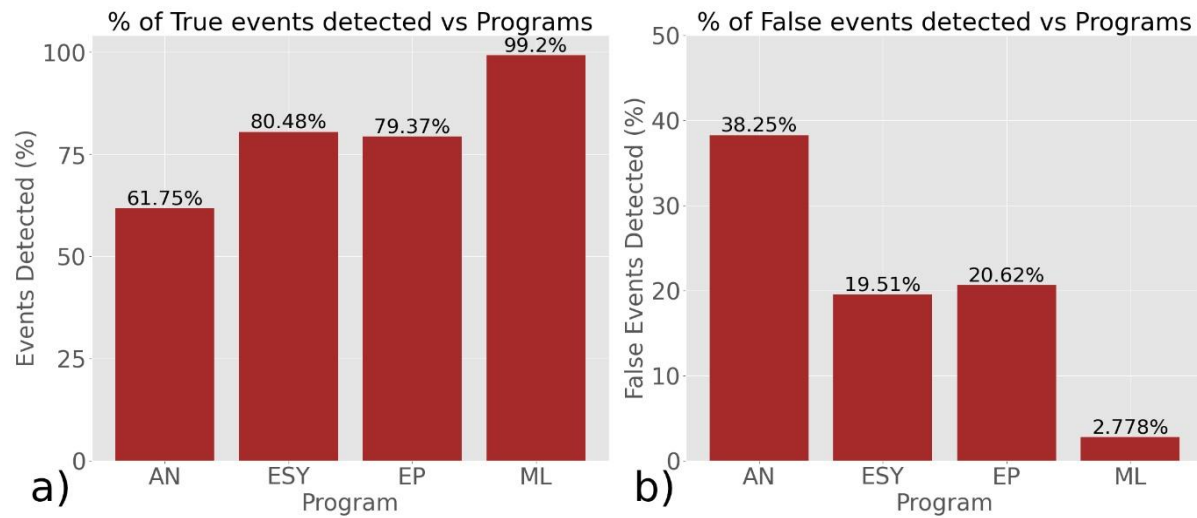

*Figure 2*: *a) Shows the percent of true detection of events by all four programs. b) Show the ratio of false detections by each program. Here AN – Autonanopore, ESY- Easynanopore, EP- EventPro and ML – Machine Learning*

5.1.2 Evaluation Time

The total processing time per file was recorded, and cumulative times were computed. EasyNanopore was the fastest, completing analysis in 12.5 seconds. EventPro and AutoNanopore took 433.7 and 447.5 seconds, respectively. The deep learning models, when tested across five random seeds, averaged 214.8 seconds. The distribution of processing times across different files is depicted in Figure 3a.

It is important to note that the times depicted in the graph represent only the runtime for each program when the correct parameters were applied to each file. The parameter tuning process through trial and error was significantly longer. Knowing the true number of events per file in the generated dataset allowed us to optimize these parameters efficiently. However, in experimental datasets, where the true number of events is unknown, parameter tuning becomes a major bottleneck and depends on the experience of a user. In contrast, the deep learning model requires no manual parameter tuning, with computation time being the only overhead for event detection.

To normalize comparisons, we scaled the runtime of each program per file by the minimum processing time and computed the average scaled processing time. EasyNanopore remained the fastest, while the deep learning models took approximately 14× longer, followed by EventPro and AutoNanopore which took 35× and 54× longer respectively. Figure 3b illustrates these results.

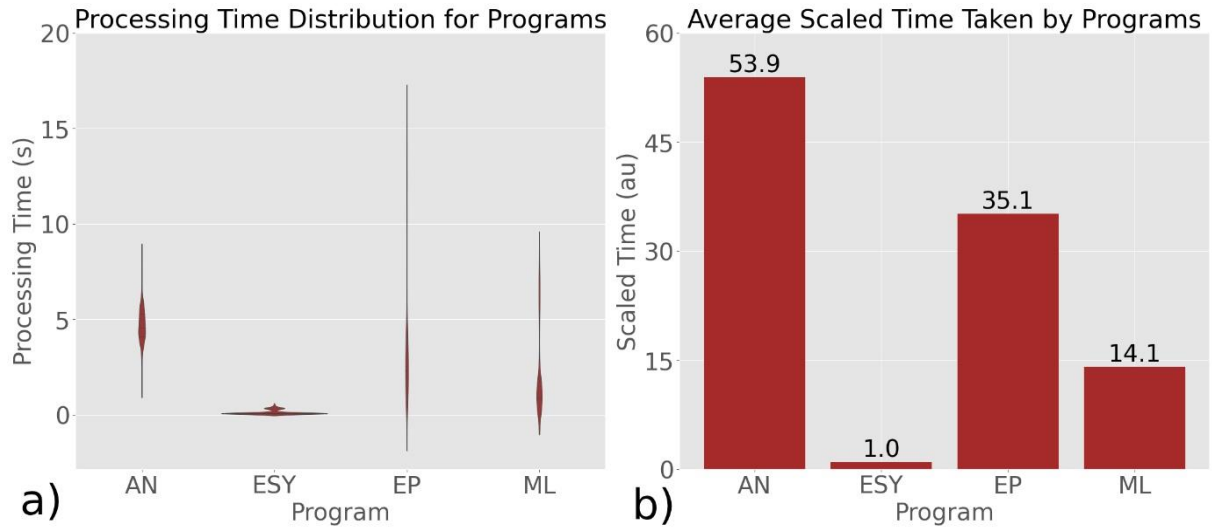

*Figure 3*: a) Illustrates the distribution of processing times for different programs. b) Shows the average scaled processing times for each program. Here AN – Autonanopore, ESY- Easynanopore, EP-EventPro and ML – Machine Learning

5.1.3 Percent Error in Width and Amplitude Estimation

We extracted the event width estimated by each program from their output data frames. The relative error in pulse width estimation was calculated for each event in each file in the dataset, culminating in the Mean Percent Error (MPE). EventPro exhibited the lowest MPE in event width estimation at 0.93%, followed by EasyNanopore (5.6%) and AutoNanopore (8.17%), as shown in Figure 4a. Similarly, we analyzed event amplitude estimation errors. EventPro again performed best, with an MPE in amplitude estimation of 10.2%, whereas AutoNanopore and EasyNanopore had significantly higher errors at 24.73% and 33.15%, respectively (Figure 4b).

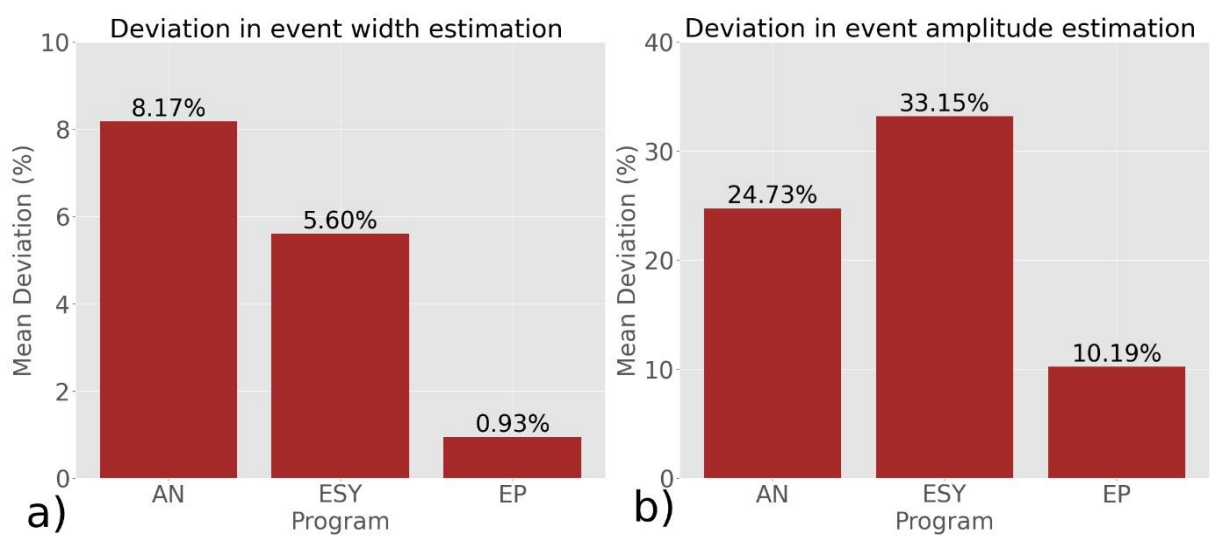

*Figure 4*: a) Illustrates average deviation in event width estimation by each program. b) Shows the average deviation in amplitude estimation. Here, AN – Autonanopore, ESY- Easynanopore, EP-EventPro

5.2 Testing on Experimental Data

For experimental data, each signal was segmented into non-overlapping 1024-point windows and analyzed using the trained model. To compare the deep learning approach with traditional methods, we benchmarked it against EventPro, which offers an intuitive GUI and event quality metrics to remove false detections/poor events. EventPro's parameters were optimized to maximize event detection and minimize false positives. Once the events were detected the poor events suggested by the program were removed. Since non-overlapping segmentation may split events across segments, we binned the detected event start and end indices into 1024-point bins to ensure a fair comparison.

The models consistently outperformed EventPro in detecting event-containing segments across all three experimental signals. Table 1 presents the highest number of event-containing bins detected for each dataset, comparing the performance of the single-channel and nine-channel models with EventPro. Since the models were trained using multiple random seeds, we evaluated the consistency of their predictions by examining the intersection of the detected bins. The overlap between both of the models was substantial, indicating a high degree of agreement, though each model also uniquely identified a few additional events. To consolidate the results, we considered the union of the predicted bins across models. Notably, each individual model outperformed EventPro, and the combined predictions further improved overall detection performance.

An interesting observation emerged from the experimental results: the nine-channel model consistently identified more high-quality events characterized by high prominence and narrow width relative to the background compared to the single-channel model. In contrast, the single-channel model was more prone to detecting lower-quality events and, though rare, occasionally produced false detections. Examples of events detected by both models across the three experimental files are provided in Section 3 of the Supporting Information.

This trend is further evident in the model performance on the experimental datasets. For instance, the *g10.abf* file featured a stable baseline with mostly narrow, high-prominence events, which the nine-channel model captured more effectively. In contrast, the other two files contained broader, multilevel events and exhibited fluctuating baselines. These differences are reflected in the number and quality of unique events detected by the single- and nine-channel models, as illustrated in the supporting figures.

We further examined the events uniquely detected by EventPro but missed by our models. These were largely found to be poor-quality detections, even after applying EventPro's built-in poor event filter. The number of bins with detected events for each program is summarized below. These findings underscore the potential of deep learning models to perform parameter-free event detection while achieving higher-quality and more consistent detection rates.

| Sl.No | File Name | Most Number of segments with events detected by DNN (9 channel) | Most Number of segments with events detected by DNN (single channel) | Number of segments with events detected by EventPro |
|---|---|---|---|---|
| 1 | G10 | 5537 | 5044 | 4950 |
| 2 | Chip A | 17168 | 18702 | 17031 |
| 3 | Chip B | 25579 | 28504 | 18728 |

*Table 1: Number of bins with events detected by model with 9 channels, single channel and EventPro.*

## 6. Results and Discussion

We developed a nanopore signal generator along with a deep learning model based on the ResNet architecture, establishing a foundation for non-parametric methods in solid-state nanopore (SSNP) signal processing. To assess the advantages of our model over classical approaches, we compared its performance with AutoNanopore, EasyNanopore, and EventPro. Using the generated dataset described in the previous section, we evaluated true and false detection rates, normalized processing times per file, and compared event width and amplitude estimation accuracy among classical software tools. Additionally, we benchmarked our model against EventPro using an experimental dataset.

This work lays the groundwork for machine learning-based, parameter-free event detection, offering a significant reduction in manual effort typically required to identify events. Remarkably, our models outperformed classical programs in detecting events from experimental datasets, despite being trained solely on a limited synthetic dataset. Performance could be further improved through training on larger datasets and by increasing model capacity. Moreover, experimental data outputs from our model can themselves be used to curate training datasets, enabling iterative refinement.

Accurate translocation event analysis depends on identifying precise start and end points, amplitudes, and multi-level structures. This can be addressed through two possible extensions. The first involves a hybrid approach: using our deep learning model to localize events, followed by classical methods to refine their properties. The second is to develop a fully end-to-end model using architectures like UNets [26], capable of direct event detection from raw signals. The modular design of our generator supports such development by enabling the creation of targeted datasets tailored to future model needs.

## Data and Code Availability

The code for the Nanopore Signal Generator GUI based on Python and will be made available in GitHub at https://github.com/jaisejohnson/NanoporeSignalGenerator. The generated dataset and an experimental signal is available upon request. Two signals used were downloaded from Drndić lab's website https://web.sas.upenn.edu/drndicgroup/data-sharing/.

## Acknowledgements

This research was supported by the Scientific and Useful Profound Research Advancement (SUPRA) Program of the Science Engineering Research Board (SERB) under Grant SPR/2021/000275.

## Author Contributions

JJ and MMV conceptualized the study. JJ developed the generator and deep learning models, generated the generated dataset, and evaluated the performance of the software under the guidance of MMV. CRG carried out the comparison and evaluation using the experimental dataset. JJ prepared the manuscript, and all authors reviewed and approved the final version.

# Supporting Information: A Solid-State Nanopore Signal Generator for Training Machine Learning Models

*Jaise Johnson[1], Chinmayi R Galigekere[2] and Manoj M Varma[1]\**

[1]*Center for Nanoscience and Engineering, Indian Institute of Science, Bangalore, Karnataka, India – 560012.*

[2]*Indian Institute of Science Education and Research, Thiruvananthapuram, Kerala, India – 695551.*

\**Email: mvarma@iisc.ac.in*

# Contents

# 1. Nanopore Data Generator

## 1.1 Introduction

The nanopore signal generator we present is a modular python program implementation aimed at addressing the lack of publicly available datasets solid state nanopore translocation experiments. The generator enables the rapid creation of large nanopore-like datasets in seconds, in contrast to the minutes or hours required for comparable experimental data collection. At the same time expensive and requires meticulous planning and designing experiments.

Corresponding to each signal, the generator produces three files:

1. A file containing the generated signal, with and without noise.
2. A file detailing the true event locations, start and end points, amplitudes, and level information for all events in the signal.
3. A file logging all generation parameters used for the corresponding signal.

These features make the generated datasets well-suited for benchmarking, as they include ground truth data that can be used to assess the accuracy of analysis tools.

Furthermore, the modular design of the generator provides granular control over noise and drift characteristics. The availability of a clean version of the signal enhances its applicability for training advanced deep learning models such as diffusion models and U-Nets.

## 1.2 Generation Process

As briefly discussed in the main text, the generation process involves the specification of various parameters to control different aspects of the signal. The generator consists of five major steps:

1. Generation of a clean signal.
2. Adding baseline drifts to the signal.
3. Applying filtering effects.
4. Addition of different noises to the clean signal.
5. Writing the final signal file, event details file and the generation parameters file.

### 1.2.1 Generating a clean signal

The gen_events function generates the clean signal based on input parameters that control the distribution and features of events. Table ST1 describes these parameters.

Event amplitudes and widths are sampled from uniform distributions using `numpy.random.uniform`, bounded by `mincurr, maxcurr, minpwd`, and `maxpwd`. Events are concatenated, and their positions are determined based on the `event_density_factor` and `dist` parameters.

.

| Sl. No | Parameter | Description |
| --- | --- | --- |
| 1 | `numpulses` | Number of events in the signal. |
| 2 | `mincurr` | Minimum amplitude of events. |
| 3 | `maxcurr` | Maximum amplitude of events. |
| 4 | `minpwd` | Minimum width of events. |
| 5 | `maxpwd` | Maximum width of events. |
| 6 | `event_density_factor` | % of datapoints of signals that should be taken up by events. |
| 7 | `dist` | Distribution of events: 'expon' (default), 'log', or 'uni' |
| 8 | `multilevel` | Whether events are single-level, multi-level, or mixed |
| 9 | `mixratio` | Ratio of multi-level events (default: 0.31) |
| 10 | `sequence` | Names for multi-levels (e.g., ['A','T','G','C']) |
| 11 | `currents` | Current values of each level |
| 12 | `pulsewidths` | Widths of each level. |
| 13 | `shuffle` | Shuffle the sequence in multi-level events. |

*Table ST 1: This table describing the different parameters in the gen_events function and the effect of each parameter on the output.*

By default, the event distribution is exponential, as biomolecular translocation is a Poisson process. After event positioning, zero arrays are added between events to construct the clean signal. The function returns the clean signal, event positions, and corresponding labels.

### 1.2.2 Adding Noise

Noise is a key feature of solid-state nanopore signals. The generator simulates noise through three distinct functions:

- White Noise: Generated from a uniform distribution scaled to 1/5th of the standard deviation of the clean signal.
- AC Noise: A 50 Hz sinusoidal waveform with up to 3 harmonics, with amplitudes scaled by $1/n^2$.
- Colored Noise (1/f noise): Generated using a power law with exponent n (default n = [1.2]).

The total noise is scaled by a factor `nsigma`, which determines the relative amplitude of events to noise.

We add all these noises together and scale this by the `nsigma`, specified in the main program. Here `nsigma` is how many times the amplitude of events is compared to the standard deviation of the noise.

### 1.2.3 Filtering effects

- RC Filtering : The function `rcdeform` models the response of a first-order RC circuit when the input signal is a current signal. It numerically integrates the governing equation of an RC circuit using a forward Euler approximation to compute the filtered response. The RC circuit is characterized by the time constant τ=RC update rule follows the discrete-time equation:

$$I_{out,i+1} = I_{out,i} + \left(\frac{\Delta t}{RC}(I_{in,i+1} - I_{out,i})\right)$$

   where Δt =1/sampling frequency, is the sampling period. The function initializes the output current to the first input value and iteratively updates it at each time step. This implementation effectively models the transient behaviour of the output current in response to a given input current signal while accounting for the filtering effect of the RC circuit.

- Low Pass filtering: implemented by the function `gaussian_low_pass_filter`, this smooths high-frequency noise and emulates the low-pass nature of experimental setups. Users can specify the cutoff frequency.

1.2.4 Adding Drifts to the Signal

In order to generate the drifts in the baseline of the nanopore current due to different factors we use two different functions.

- Sinusoidal Drifts (`sinusoidal_drifter`): This function takes in the generated signal and two parameters to add sinusoidal drifts.
   - `numconcs` : Number of sinusoids concatenated.
   - `maxamp` – Maximum amplitude of the sinusoids.

Amplitudes and harmonics are randomly chosen within the specified bounds.

- Abrupt Drifts (`abrupt_drifter`):
- `nstepwins`: Number of step segments
- `driftmaxmag`: Maximum drift amplitude
- `maxnsteps`: Maximum number of sublevels per step

Both drift types simulate physical changes like pore shrinkage or abrupt surface changes.

These baseline drifts are added to the generated signal to achieve a drifted signal as a final output. Thus, the generator enables the generation of different type of signals this is illustrated in the figure SF1 below.

1.2.5 Writing the final signal files along with event details and generation parameters

All components are integrated in the final function. The SNR is controlled using nsigma via three strategies:

1. Minimum amplitude-based scaling (default): Ensures all events exceed nsigma * std(noise)

2. Maximum amplitude-based scaling: Ensures events do not exceed nsigma * std(noise)

3. Average amplitude-based scaling: Uses mean amplitude for SNR matching

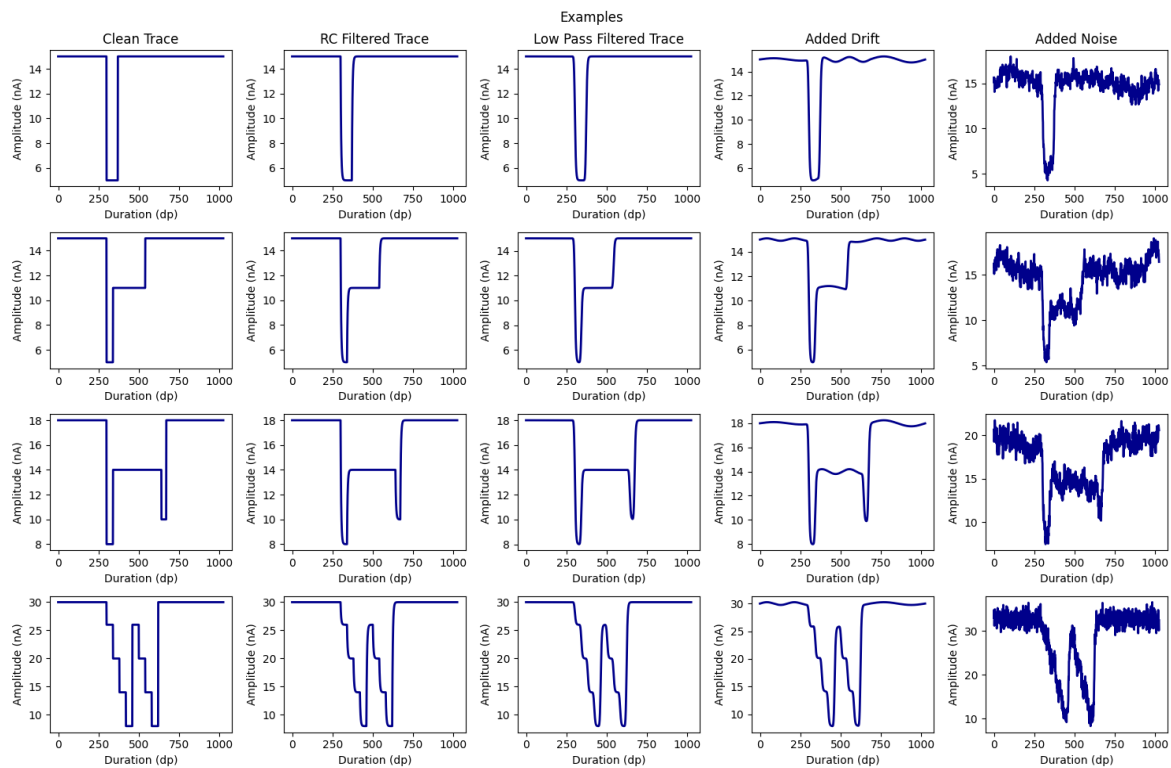

*Supporting Figure SF 1: Different types of signals generated at various steps of generation.*

By default, the first strategy is used, but the users can modify the scaling factor in the code if a different method is preferred.

After scaling, RC filtering and low-pass filtering are applied. The filtered signal is combined with noise and drift to form the final signal. A `vshift` parameter can be applied to adjust the open-pore baseline current, reflecting experimental variability. This can be tuned to adjust the prominence of events.

Once the final signal is generated, data is stored as a pandas dataframe. The `sampfreq` parameter, allows users to specify the sampling rate, and based on this and number of datapoints a corresponding 'Time' array is added to the dataframe. Users can choose to include additional components such as the clean signal, filtered signal, drift signal, and noise to this dataframe, leveraging the modularity of the program. This dataframe is saved as a .csv file upon the completion of the generation run.

Additionally, an *event details file* (.csv file) is generated, containing key information such as event start points, end points, event widths, mean current levels, level widths, and event amplitude values. A separate *parameter log* (.txt file) is also created, ensuring that users can easily track the generation parameters used for each generated signal file.

## 1.3 Generator GUI

To make the generator accessible to users without programming expertise, a GUI is provided. Users can modify all parameters through the interface and view a plot of the generated signal upon completion.

- Tab 1: Signal generation parameters and waveform plot
- Tab 2: Power spectral density (PSD) of the signal
- Tab 3: Event details table (corresponds to the event .csv file)

This organization allows users to assess results and perform new generation runs without leaving the GUI.

Supporting Figures SF2–SF4 show screenshots of the GUI tabs.

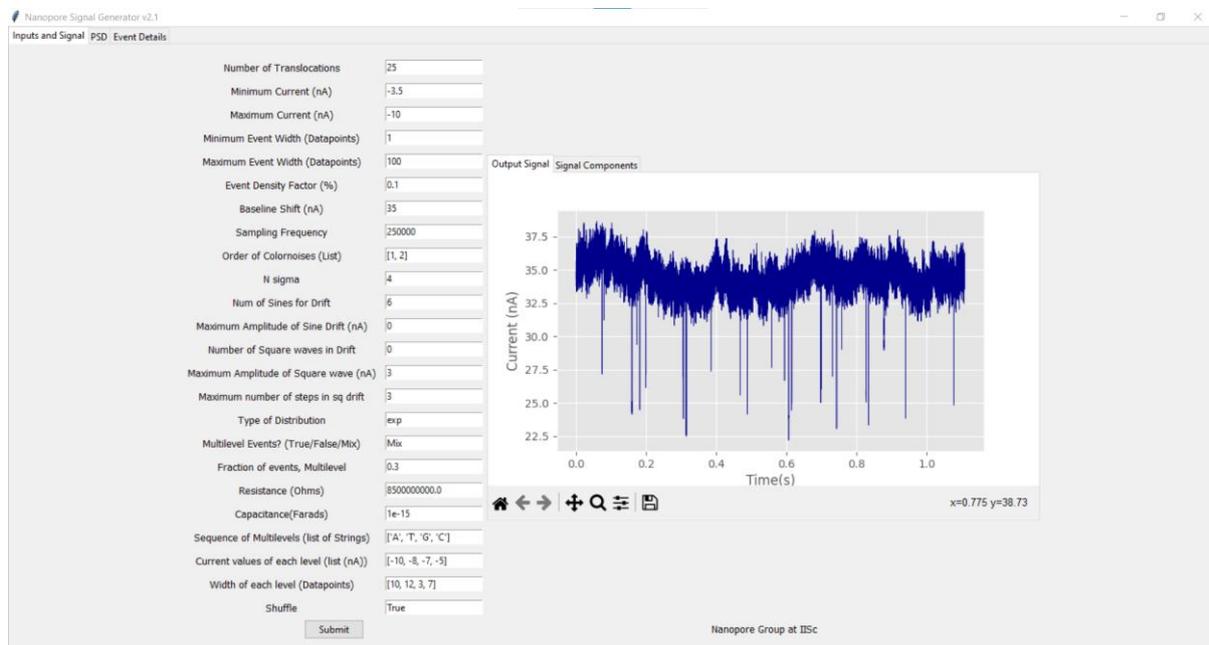

*Supporting Figure SF 2: The GUI window of the generator after a successful run.*

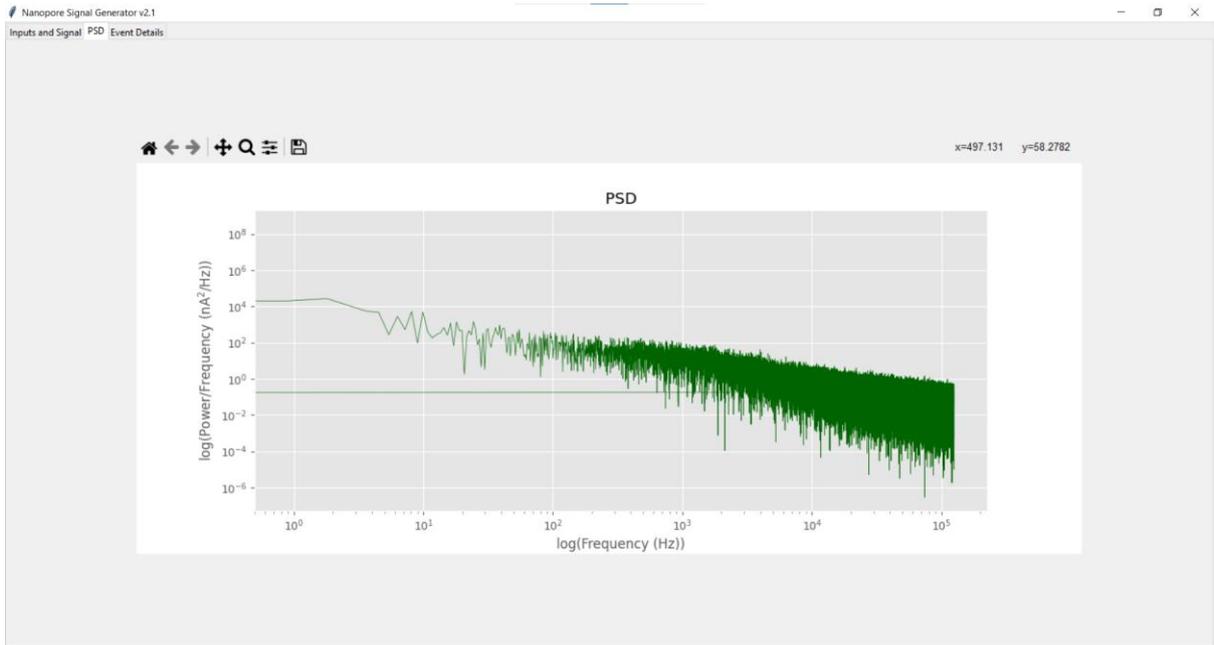

*Supporting Figure SF 3: The PSD tab of the GUI window of the generator after a successful run.*

*Supporting Figure SF 4: The Event Details tab of the GUI window of the generator after a run.*

# 2. Training a CNN for event detection using generated dataset.

## 2.1 Introduction

Use of machine learning methods will ease data analysis of nanopore current data. The potential of unlocking parameter free event detection holds great potential in the rapid adoptability of nanopore devices for wide variety of applications.

## 2.2 Dataset

A dataset for training and testing the Deep learning model was generated by generator by modifying the program to randomly initialize different input values within a specified range and iterating. We generated a dataset with signals of each 1024 datapoints long containing different types of events and baseline currents by this method to finally have a dataset with 40,900 instances which was further used for training and testing of the model. Fig SF5 shows a few examples from the generated dataset.

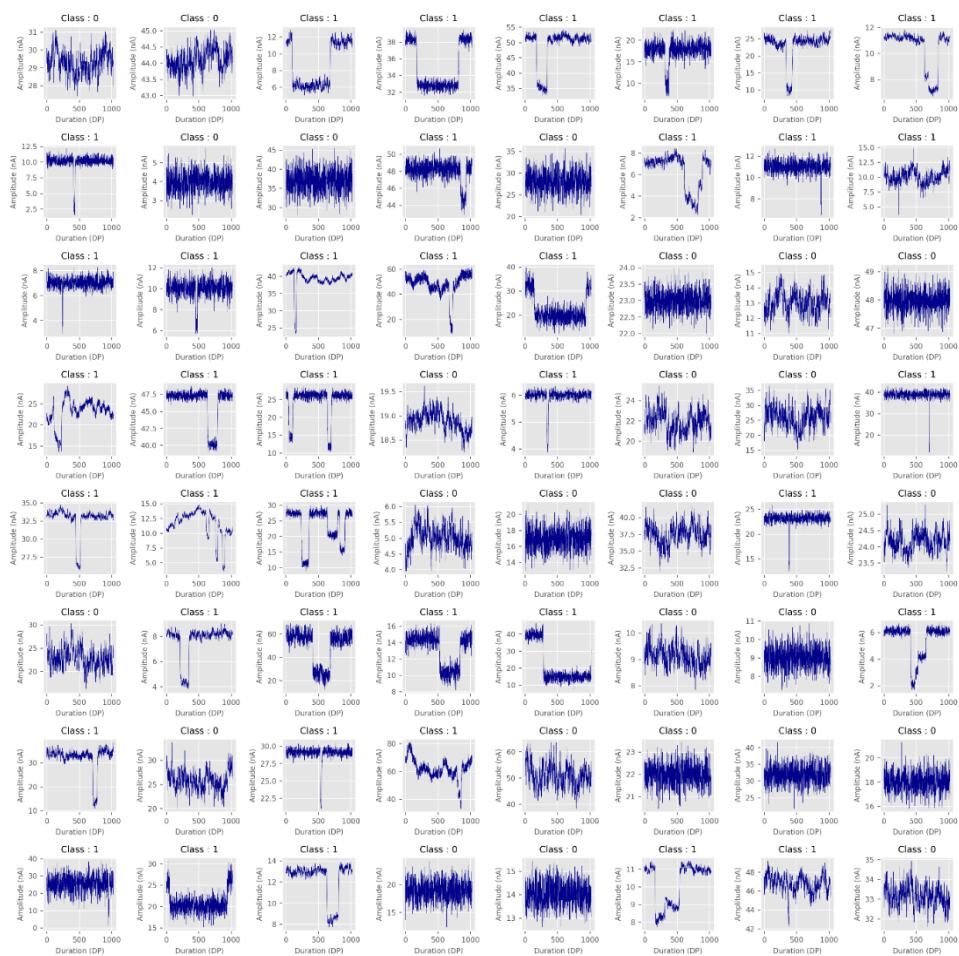

*Supporting Figure SF 5: Example batch of training data, the title 'Class: n' where n = 0/1 represents the label assigned to each signal. (1 represents signal with event, 0 represents signal without event)*

## 2.3 Model architecture

The structure of the 9 channel model is described below. The single-channel model shares the same architecture, except for the input layer, which accepts a single channel instead of nine.

The model is a deep 1D convolutional network designed for nanopore signal processing, consisting of 14 trainable layers, including 5 convolutional blocks, 3 residual blocks, a bypass max-pooling pathway, and a fully connected classifier. The convolutional blocks apply a 1D convolution (kernel size 33, stride 1, padding 16), followed by Instance Normalization, LeakyReLU activation, and Dropout (20%), with optional max-pooling (kernel size 2). The residual blocks consist of two consecutive 1D convolutions (kernel size 3, stride 1, padding 1), each followed by Instance Normalization, LeakyReLU activation, and Dropout, with a skip connection ensuring stable gradient flow.

The input signal (9 channels) first passes through a convolutional block (Conv1) to expand the feature dimension to 16 channels. Subsequent conv_blocks (Conv2 and Conv3) progressively increase the feature depth to 32 and 128 channels, interleaved with residual blocks for enhanced feature extraction. After further refinement through Conv4 (64 channels) and Conv5 (16 channels), a bypass max-pooling pathway processes the early feature maps in parallel, merging them with the final extracted features. The classifier applies max-pooling, dropout, a fully connected layer, LeakyReLU, and a final sigmoid activation to produce the classification probability.

*Supporting Figure SF 6: nine- channel model architecture*

## 2.4 Data Preprocessing

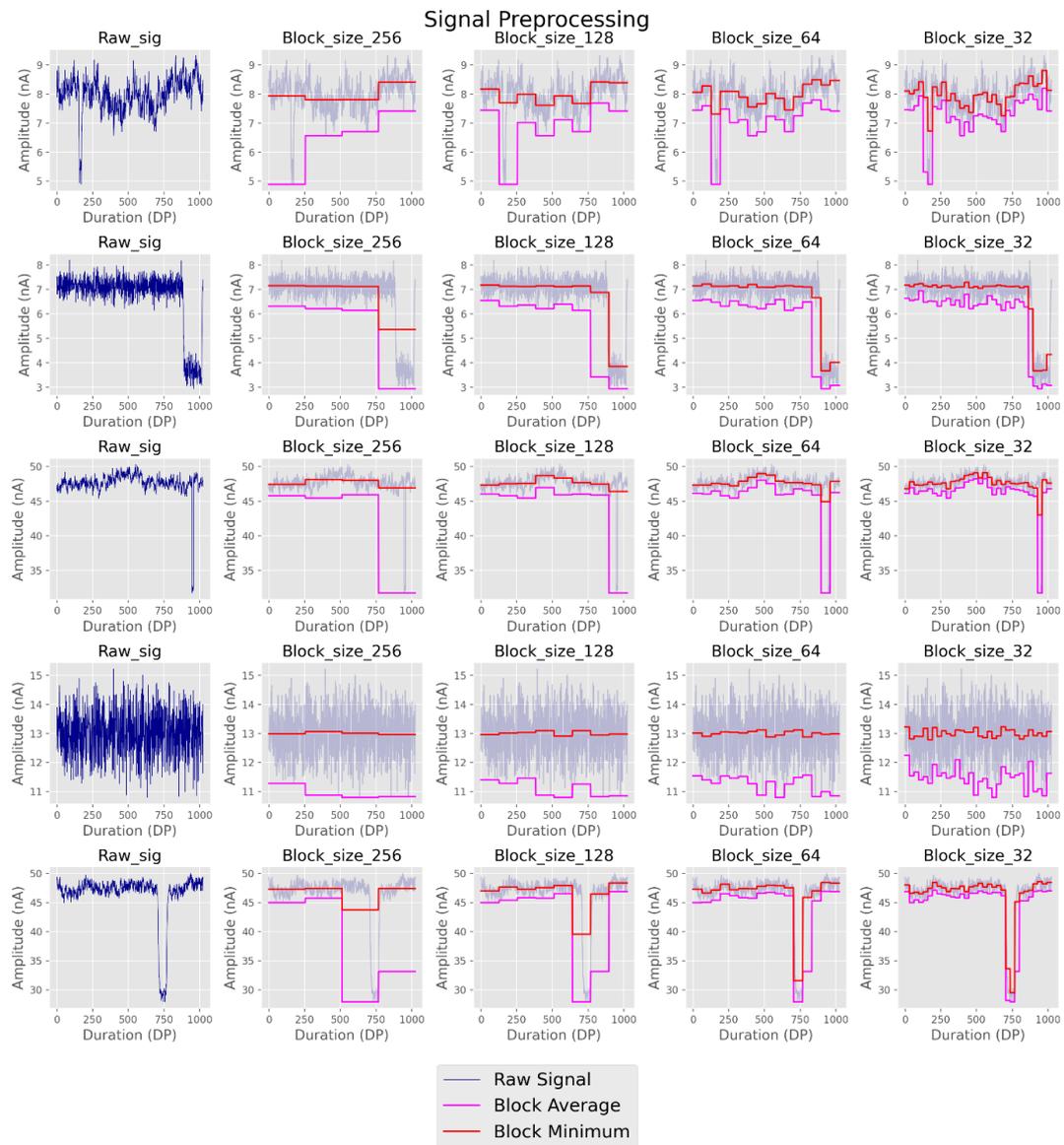

*Supporting Figure SF 7: Block averaging and block minimization to enhance the input signal before passing in to the model.*

## 2.5 Training

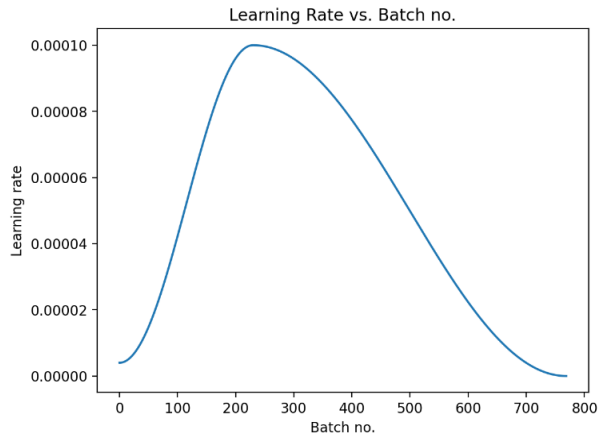

*Supporting Figure SF 8: Learning rate adjustment with batches.*

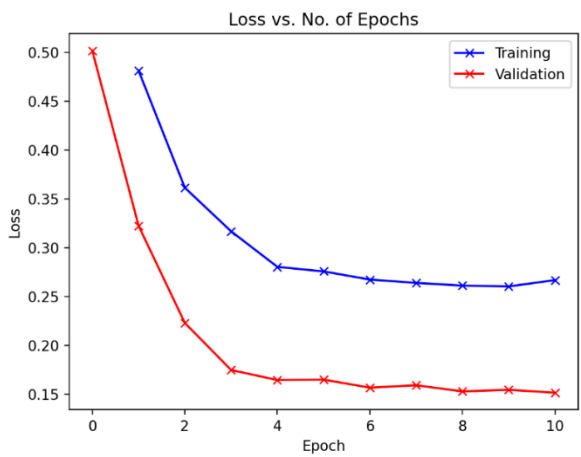

*Supporting Figure SF 9: Training and validation loss with epochs.*

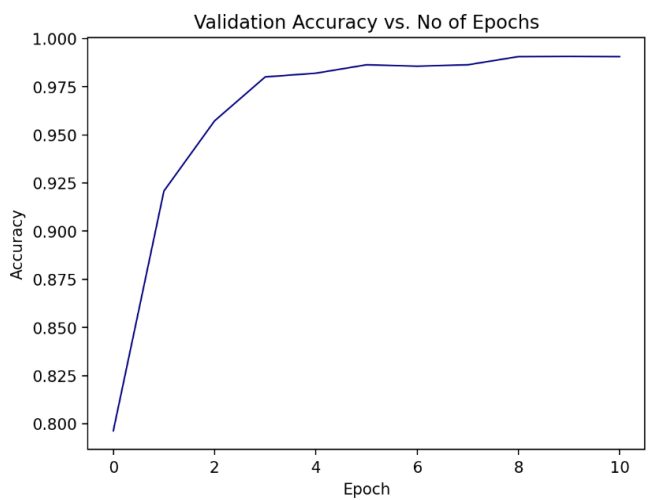

*Supporting Figure SF 10: Improvement in validation accuracy with epochs.*

# 3. Generated Dataset for Benchmarking

## 3.1 Introduction

To demonstrate the utility of the generated dataset in benchmarking nanopore signal analysis software platforms we conducted a comprehensive evaluation of them on a generated dataset as mentioned in the main article. Here we provide a detailed description of the dataset along with the rationale behind the chosen distribution of parameters in the benchmarking dataset.

The dataset consists of 105 generated signals, evenly distributed to 5 different classes of different event density factors: 0.1, 0.5, 1, 5, and 10. This event-density factor represents the approximate percent of data points in a signal that corresponds to translocation events.

That is if the event-density factor is set at 10, the total length $l$, of the generated current trace is estimated as

$$l = \frac{n \times 100}{10}$$

Where, $l$, is the total length of the current trace.

$n$, is the number of datapoints taken up by all events.

We vary the number event density factor in the dataset to account the different trends of data observed due to difference in experimental setups. The appearance of translocation events in an experiment is influenced by multiple factors, including applied voltage, pore diameter, analyte cross-section, electrode material, membrane charge distribution, pore structure, and wettability, among others. Adjusting the event-density factor allows us to simulate systemic changes in data while maintaining control over the event distribution within a signal.

Within each event-density factor there are 21 files with different attributes. We have three different noise levels, namely 3, 5, 7 times the standard deviation of noise floor. Corresponding to each noise level we have seven different files with different `vshift` values.

To ensure a consistent noise-level factor across all events within a signal and to enable reliable benchmarking, the event amplitude in all signals was fixed at $-3$ nA. Noise levels were adjusted accordingly, as described earlier.

Maintaining a constant event amplitude across all signals mimic real nanopore translocation experiments where the same analyte is measured across different experimental conditions by altering factors such as pore diameter and material composition. This approach enables us to construct a dataset optimized for benchmarking purposes.

For signals with event-density factor = 10, event widths were varied in the range [100, 1000] to ensure that the overall signal length remained sufficient for analysis using EventPro. For all other sparsity levels, event widths (analogous to dwell times in real translocation experiments) followed a uniform distribution in the range [10, 100] data points. Figure SF11 below illustrates the distribution of events in the generated dataset.

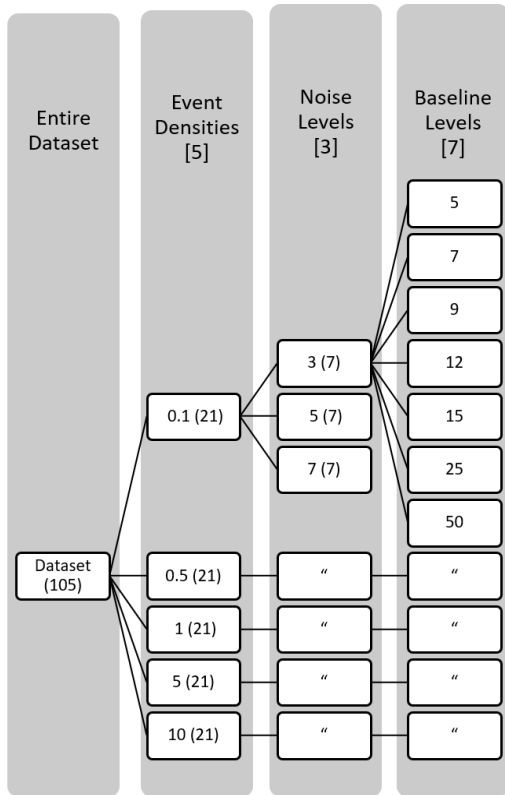

*Supporting Figure SF 11: Illustration of distribution of events based on generation parameters.*

# 4. Results

The trained model performed well on the test dataset, derived from the original dataset and achieved an accuracy of 99%.

| Sl.No | Seed | Model (9 Channel) False Detection (%) | Model (1 Channel) False Detection (%) | Model (9 Channel) True Detection (%) | Model (1 Channel) True Detection (%) |
|---|---|---|---|---|---|
| 1 | 13 | 1.633141 | 4.137925 | 99.1563 | 99.2708 |
| 2 | 37 | 2.004229 | 3.589345 | 98.9792 | 99.4687 |
| 3 | 23 | 2.631090 | 3.392576 | 98.2604 | 99.4687 |
| 4 | 57 | 1.763557 | 3.605953 | 99.1458 | 99.5104 |
| 5 | 42 | 3.478010 | 2.407348 | 98.6875 | 99.5625 |

*Table ST 2: This table shows true and false detection of events from the generated test dataset for models trained on different distribution of train, validation and test dataset using different random seeds.*

As mentioned in the main text, we tested the performance of the model on datasets derived from real experiments. Table ST 2 shows the results of training on different distributions random distributions of the dataset and testing on experimental data.

Figures SF 12 - SF 29 shows examples of detected event bins and their indices for each of the files by models and EventPro along with the unique events detected by the models combined and EventPro.

| Sl.No | File Name | Random Seed | Number of bins with events detected by DNN (9 channel) | Number of bins with events detected by DNN (single channel) | Number of segments with events detected by EventPro | Number of bins uniquely identified by eventpro and not by ML models. | Number of bins uniquely identified by ML and not by eventpro. |
|---|---|---|---|---|---|---|---|
| 1 | G10 | 42 | 5368 | 4858 | 4950 | 354 | 1321 |
|   |   | 13 | 5493 | 4873 |   | 334 | 1393 |
|   |   | 37 | 5369 | 5044 |   | 354 | 1369 |
|   |   | 23 | 5303 | 4862 |   | 369 | 1373 |
|   |   | 57 | 5537 | 4810 |   | 343 | 1411 |
| 2 | Chip A | 42 | 16003 | 17882 | 17031 | 1313 | 5801 |
|   |   | 13 | 17168 | 18284 |   | 1262 | 6934 |
|   |   | 37 | 16558 | 18702 |   | 1327 | 6871 |
|   |   | 23 | 15790 | 16503 |   | 1485 | 4503 |
|   |   | 57 | 16053 | 17496 |   | 1377 | 5561 |

| 3 | Chip B | 42 | 20595 | 26842 | 18728 | 812 | 14962 |
|   |        | 13 | 25579 | 27497 |       | 628 | 17610 |
|   |        | 37 | 23320 | 28504 |       | 755 | 17199 |
|   |        | 23 | 18808 | 24324 |       | 1124 | 12341 |
|   |        | 57 | 20495 | 26455 |       | 804 | 16300 |

*Table ST 3: This table shows the number of events detected by the models trained on different distribution of train, validation and test dataset using different random seeds on experimental test data.*

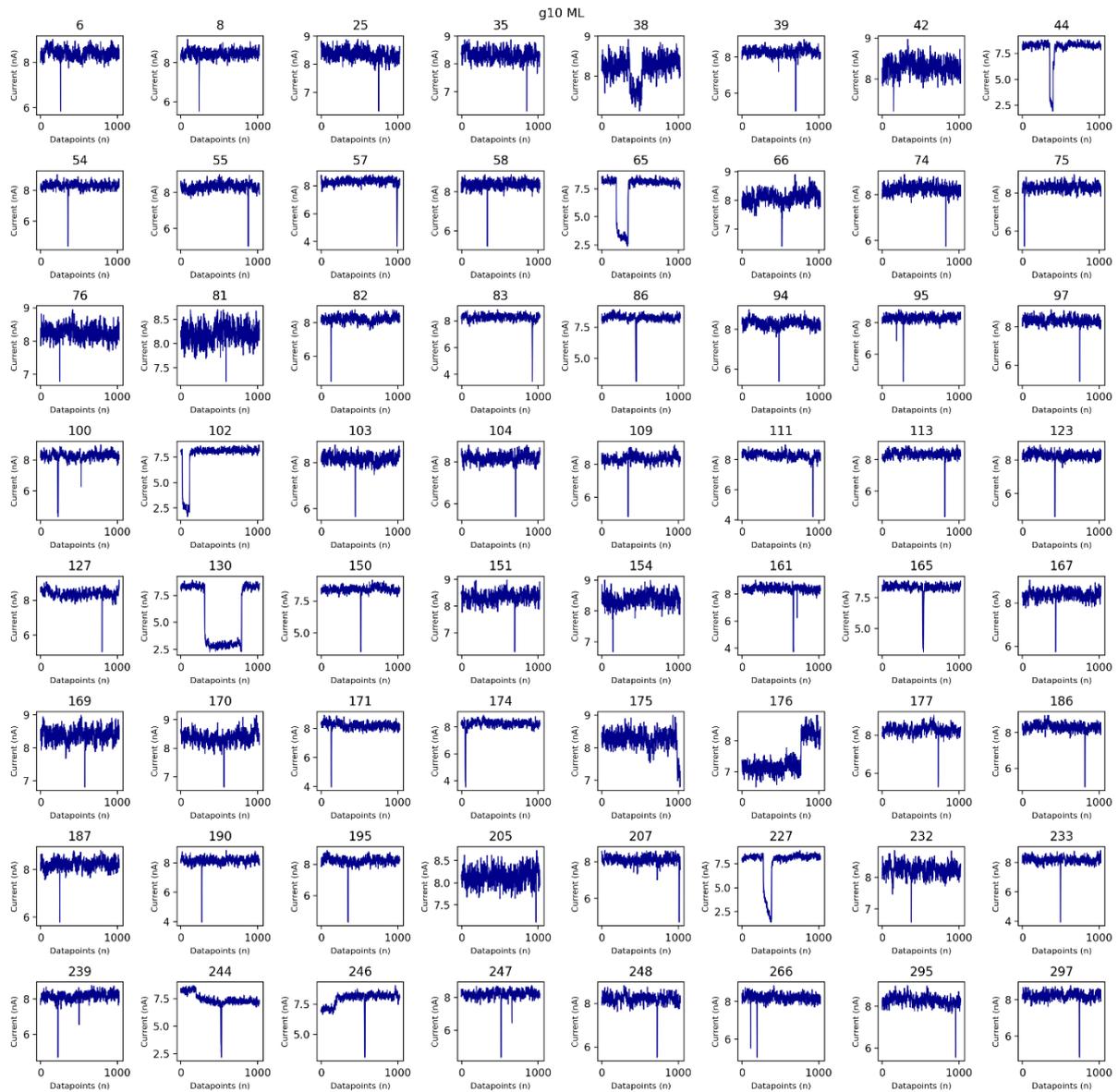

*Supporting Figure SF 12: Few event bins detected by the models (Union of events detected by single channel and 9 channel models) for the file g10.abf, the title of each figure represents the index of the bin.*

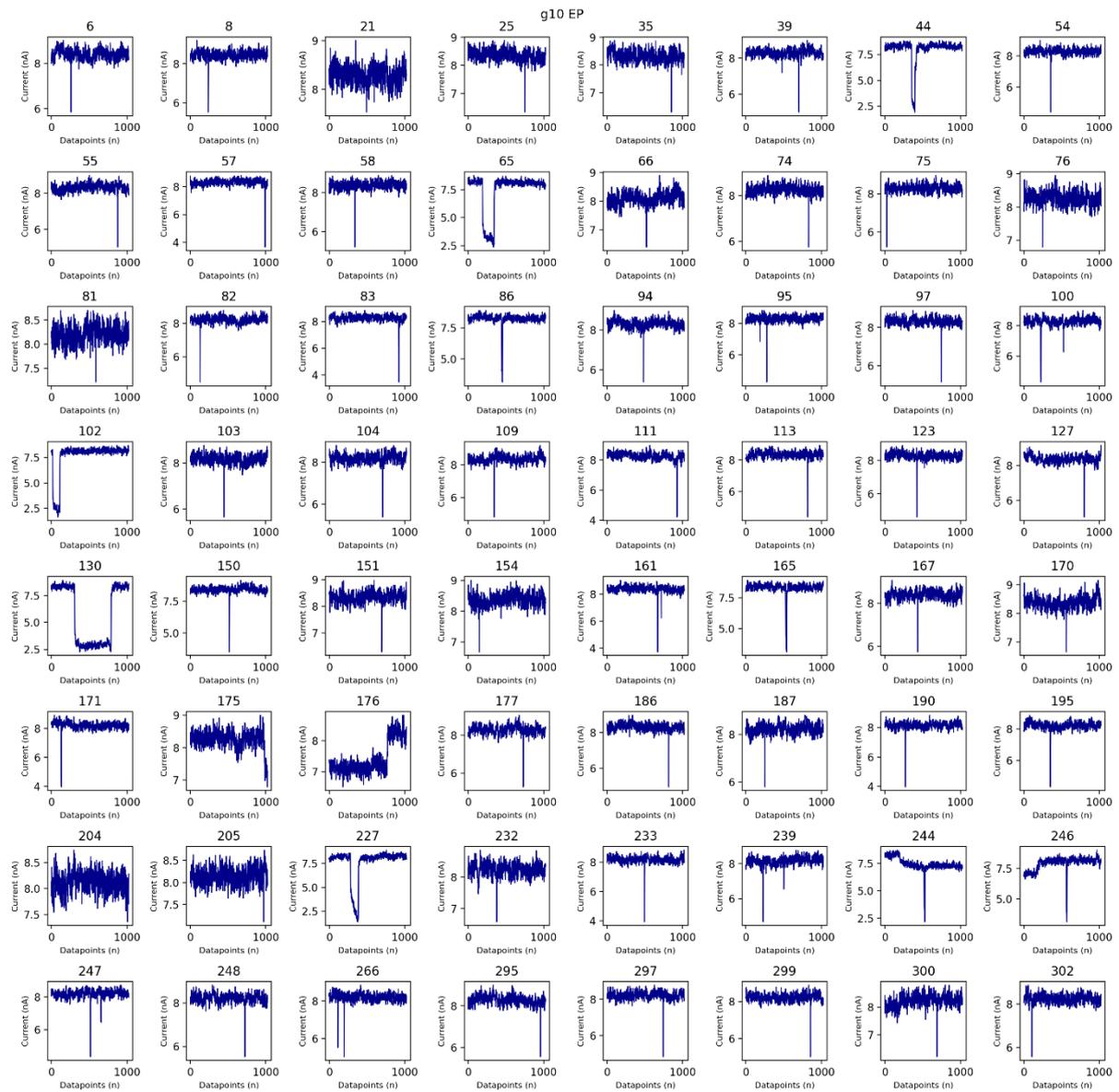

Supporting Figure SF 13: Few event bins detected by EventPro for the file g10.abf, the title of each figure represents the index of the bin.

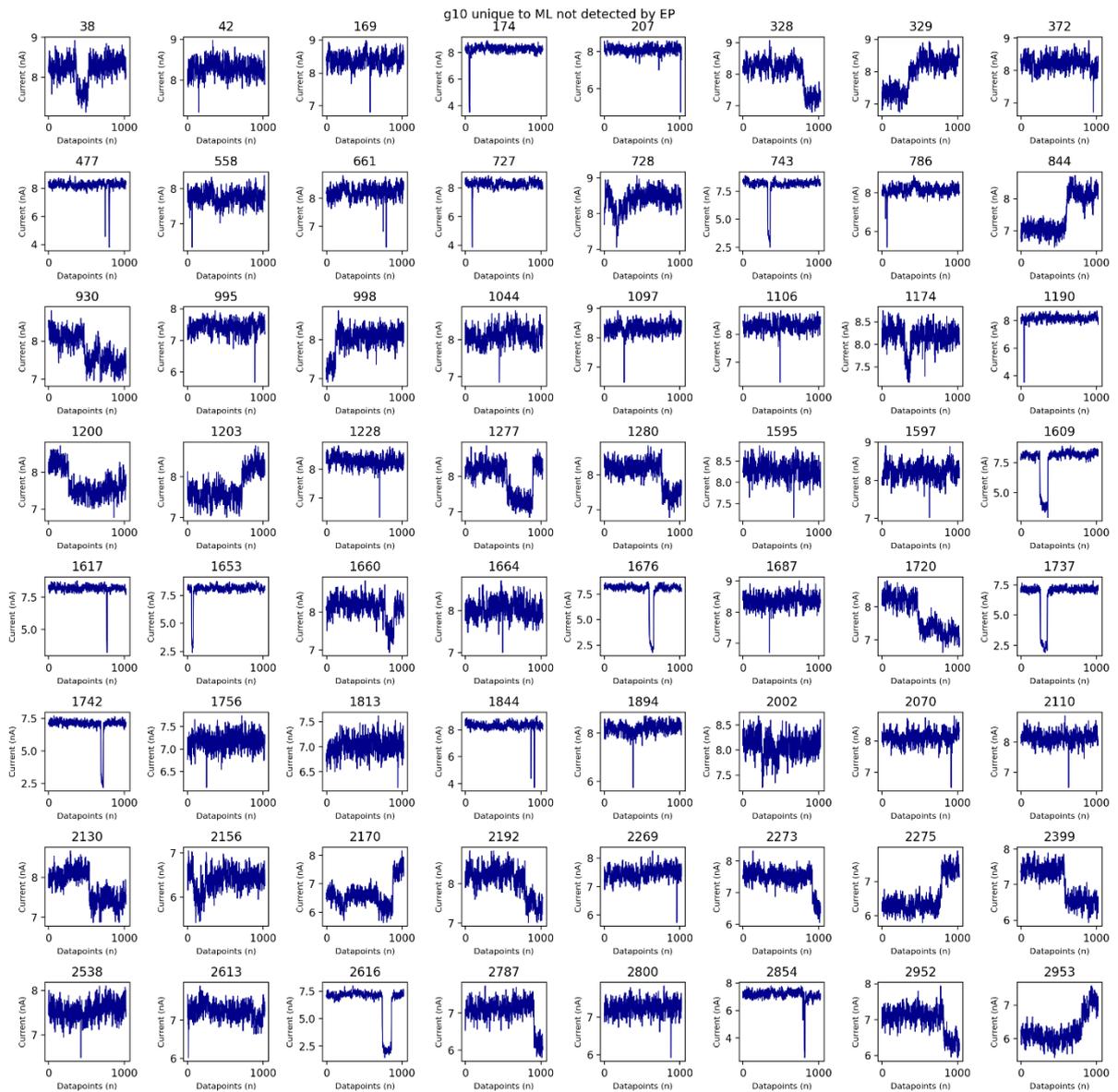

*Supporting Figure SF 14: Few event bins detected by the models but not detected by EventPro for the file g10.abf, the title of each figure represents the index of the bin.*

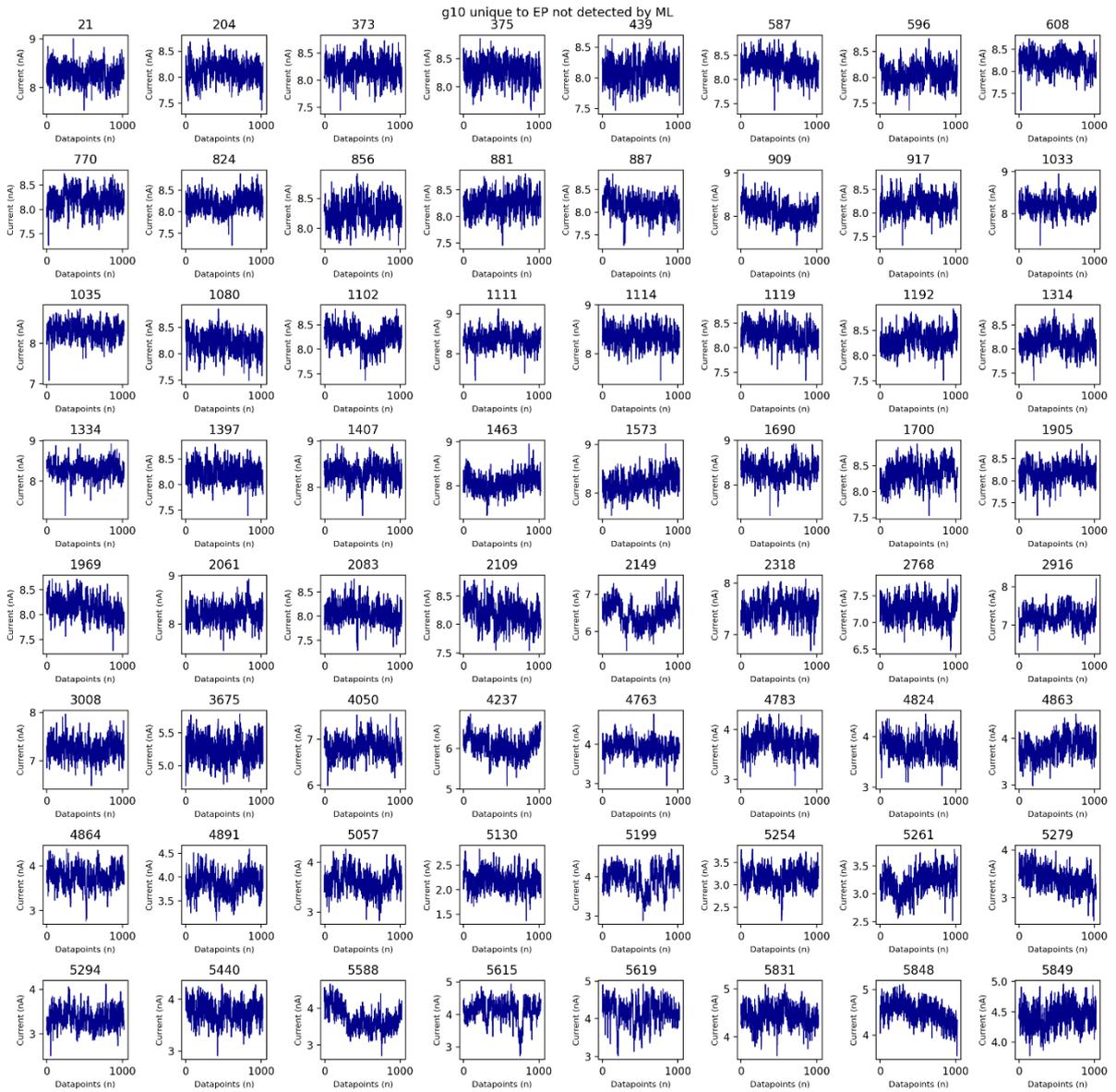

*Supporting Figure SF 15: Few event bins reported as events by EventPro, but undetected by the models for the file g10.abf, the title of each figure represents the index of the bin.*

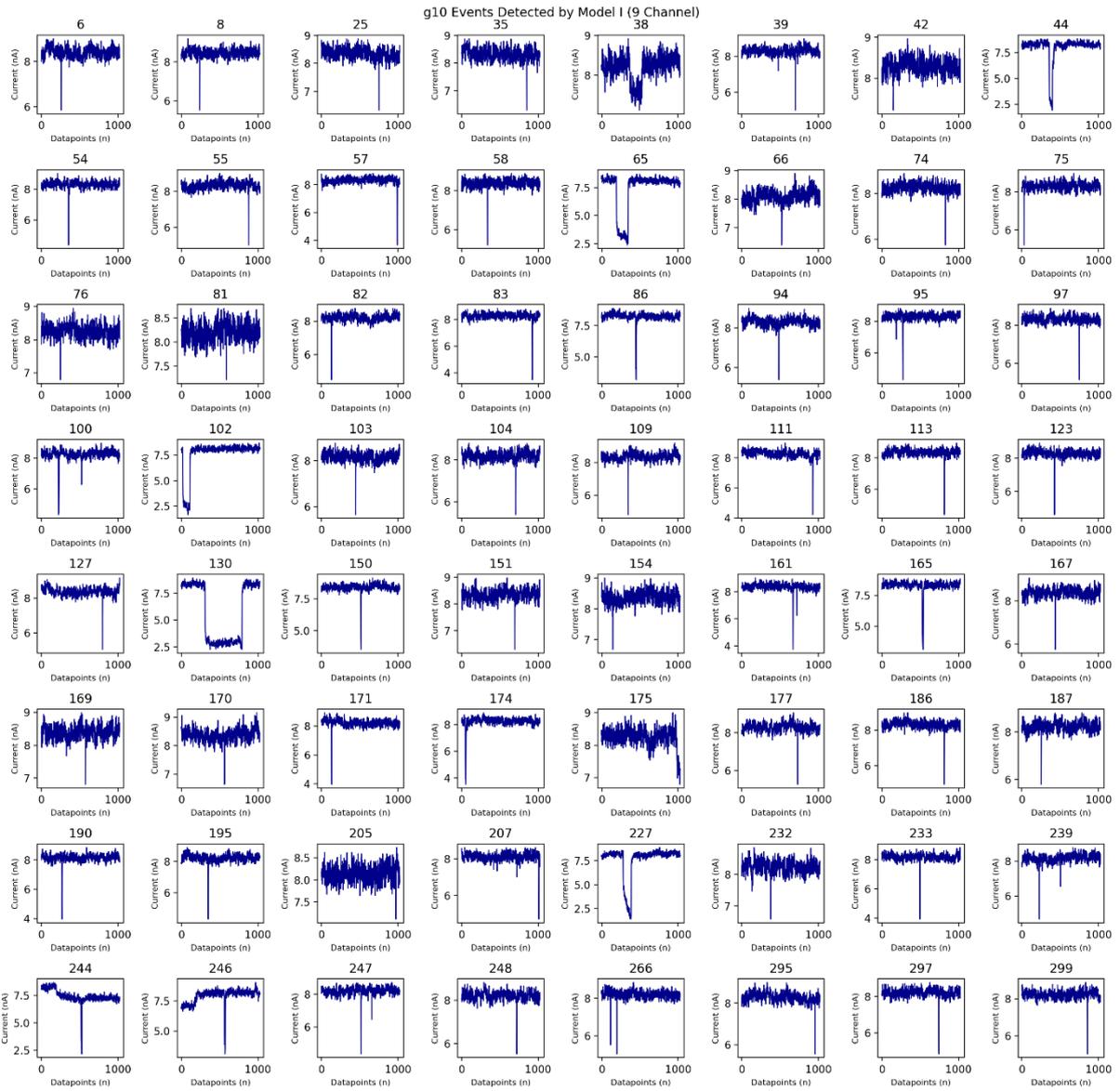

*Supporting Figure SF 16: Few event bins detected by the 9-channel model from g10.abf file.*

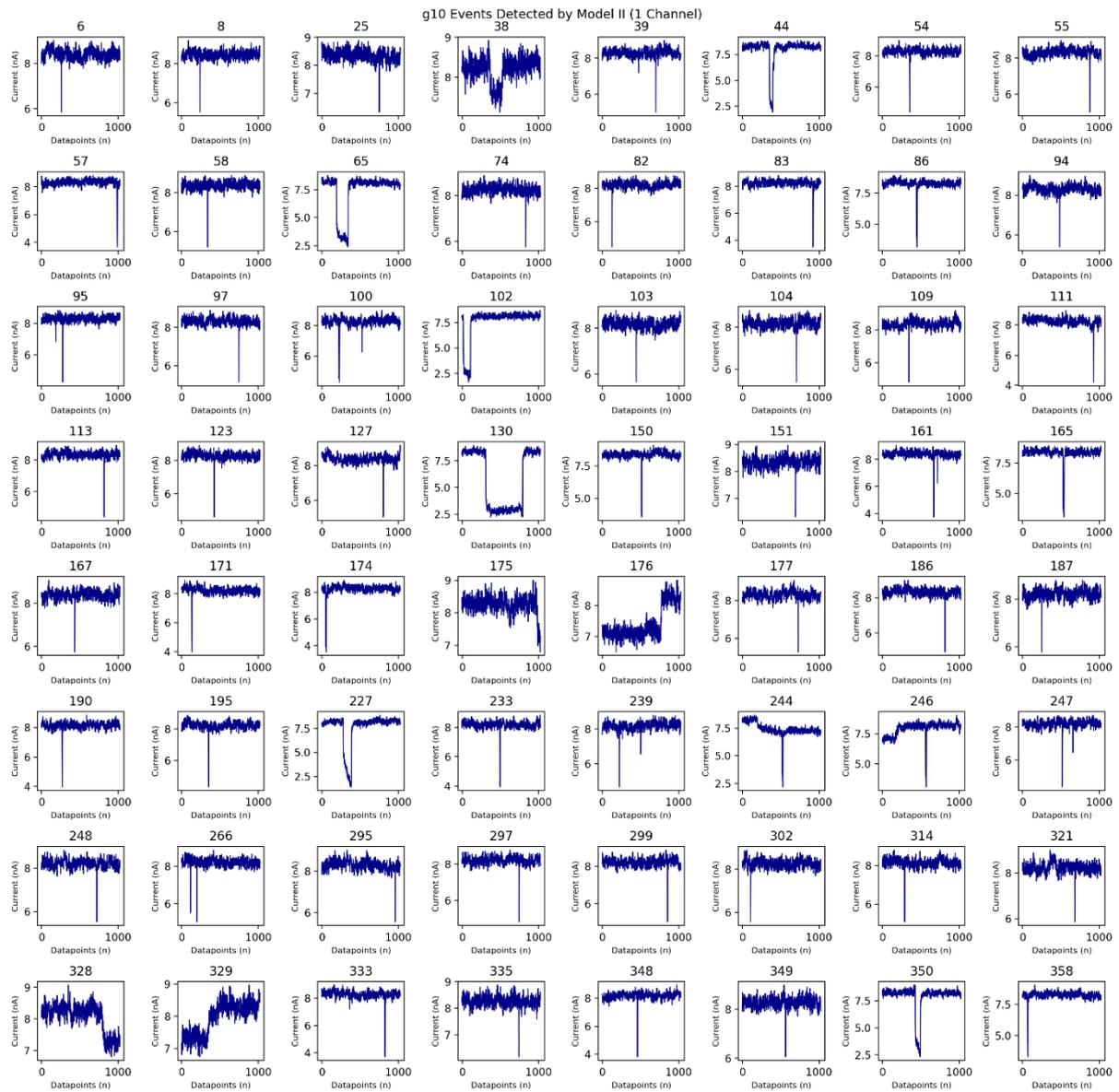

*Supporting Figure SF 17: Few event bins detected by the single channel model from g10.abf file.*

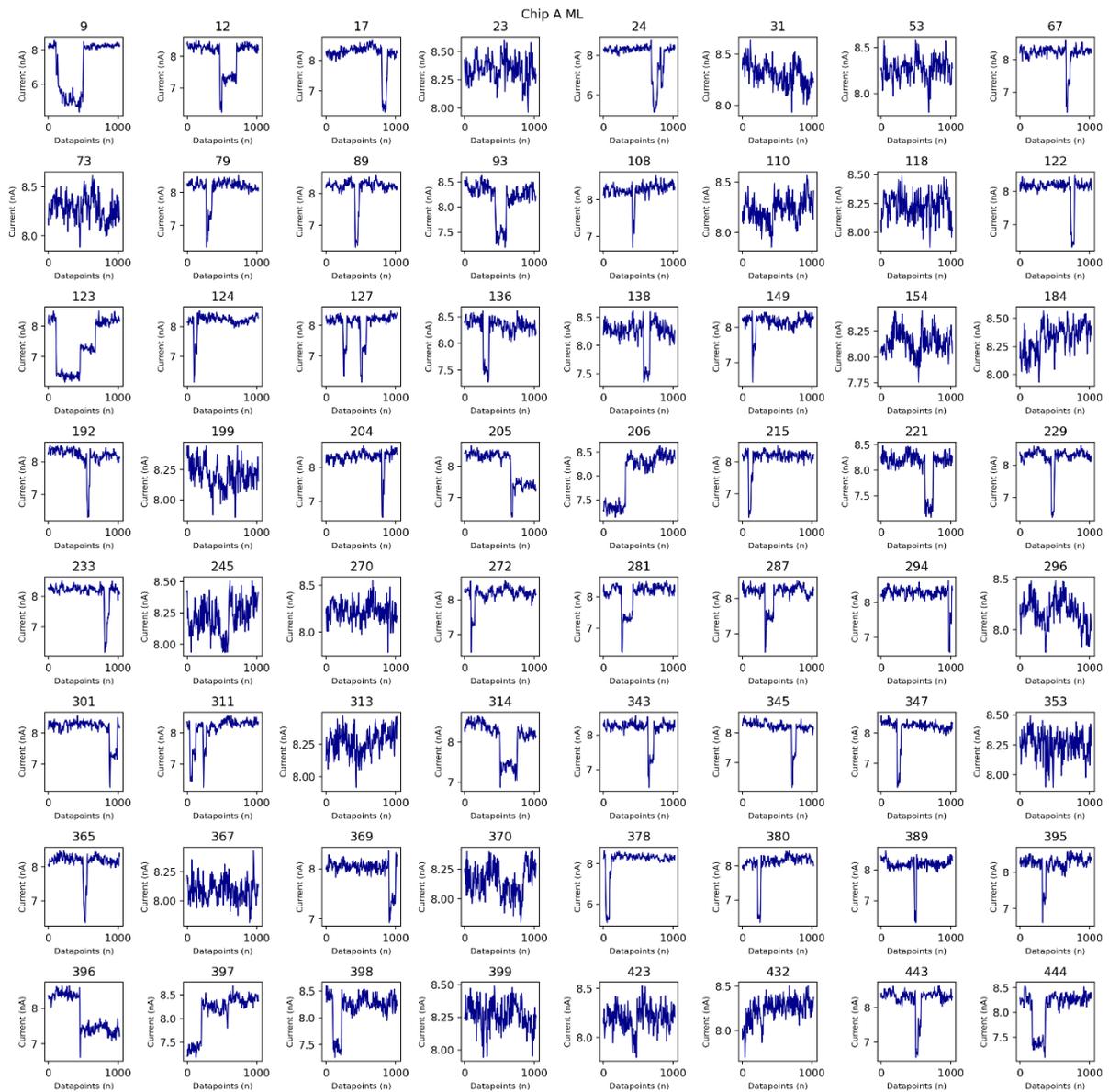

Supporting Figure SF 18: Few event bins detected by the models (Union of events detected by single channel and 9 channel models) for the file ChipA.abf, the title of each figure represents the index of the bin.

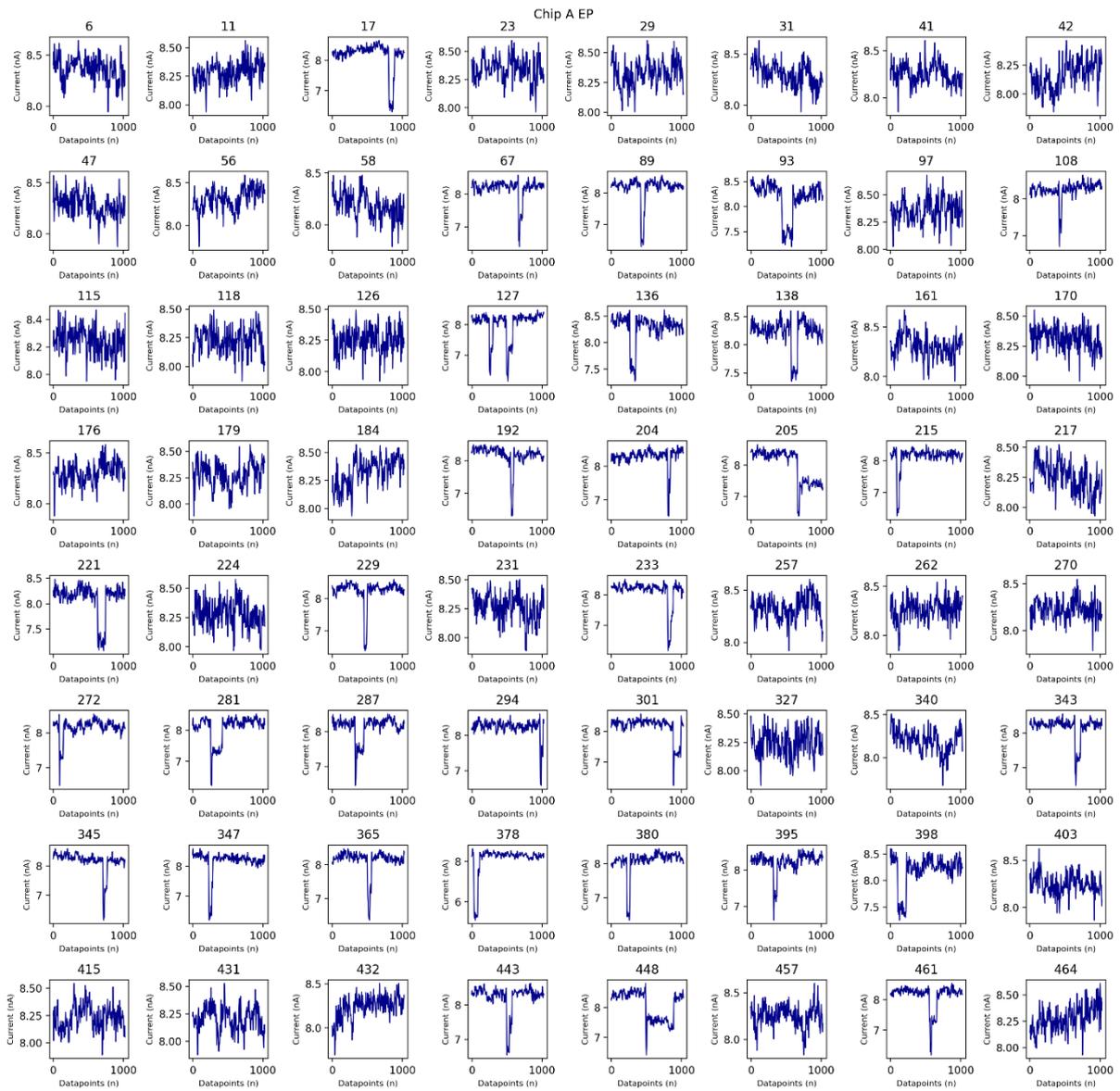

*Supporting Figure SF 19: Few event bins detected by the models but not detected by EventPro for the file chipA.abf, the title of each figure represents the index of the bin.*

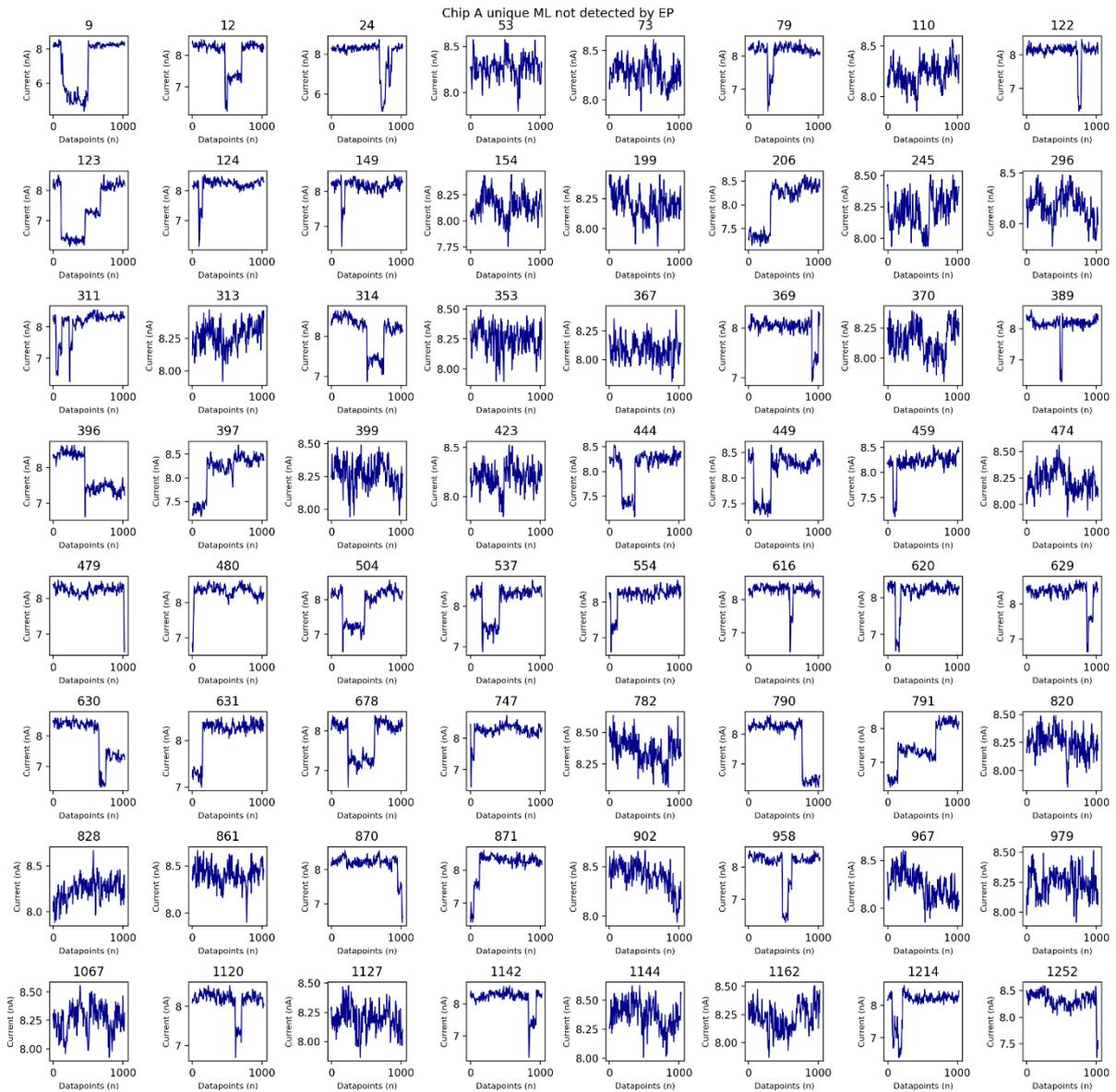

Supporting Figure SF 20: Few event bins detected by the models(Union of events detected by single channel and 9 channel models) but not detected by EventPro for the file chipA.abf, the title of each figure represents the index of the bin.

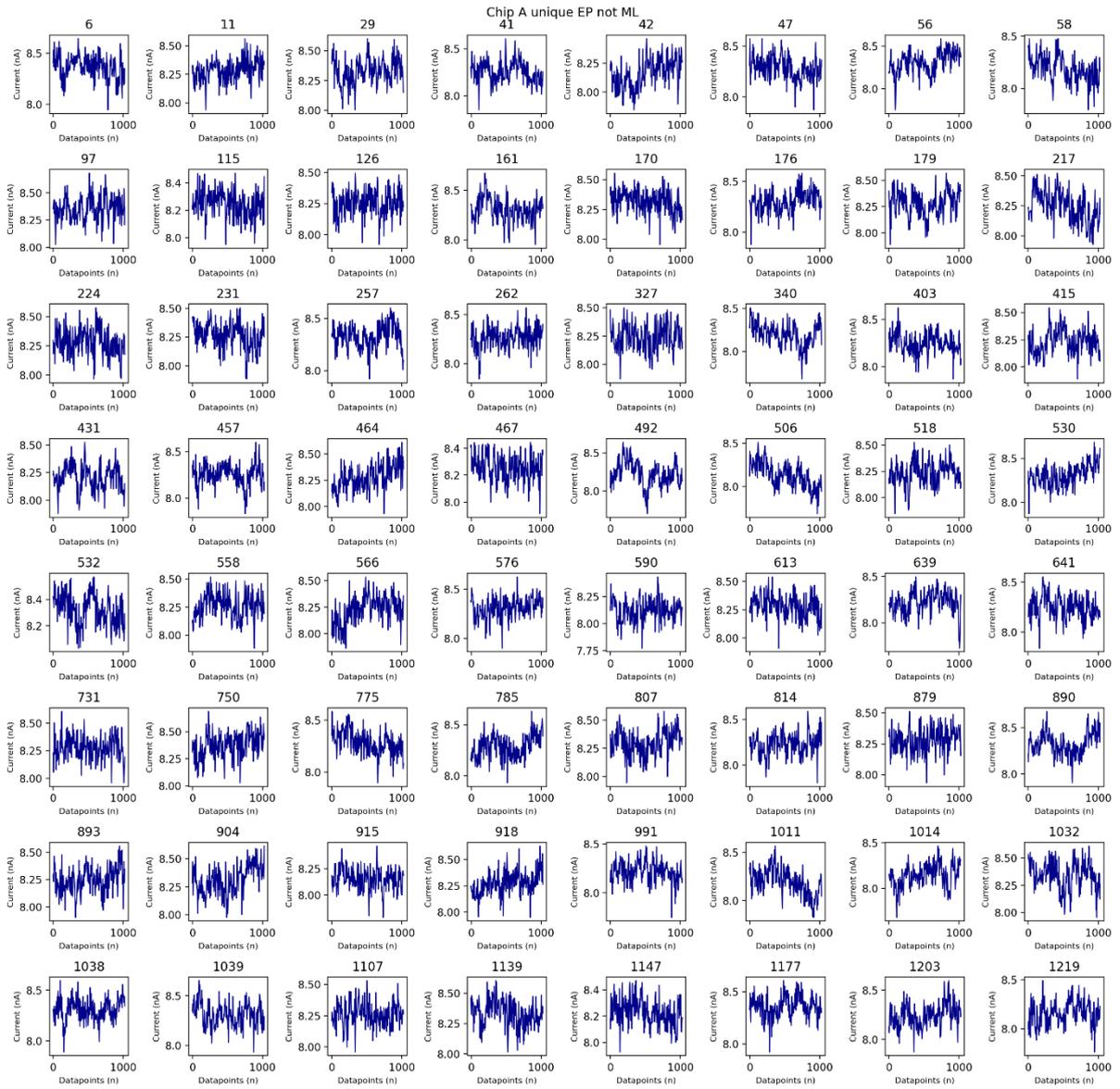

*Supporting Figure SF 21: Few event bins reported as events by EventPro, but undetected by the models for the file ChipA.abf, the title of each figure represents the index of the bin.*

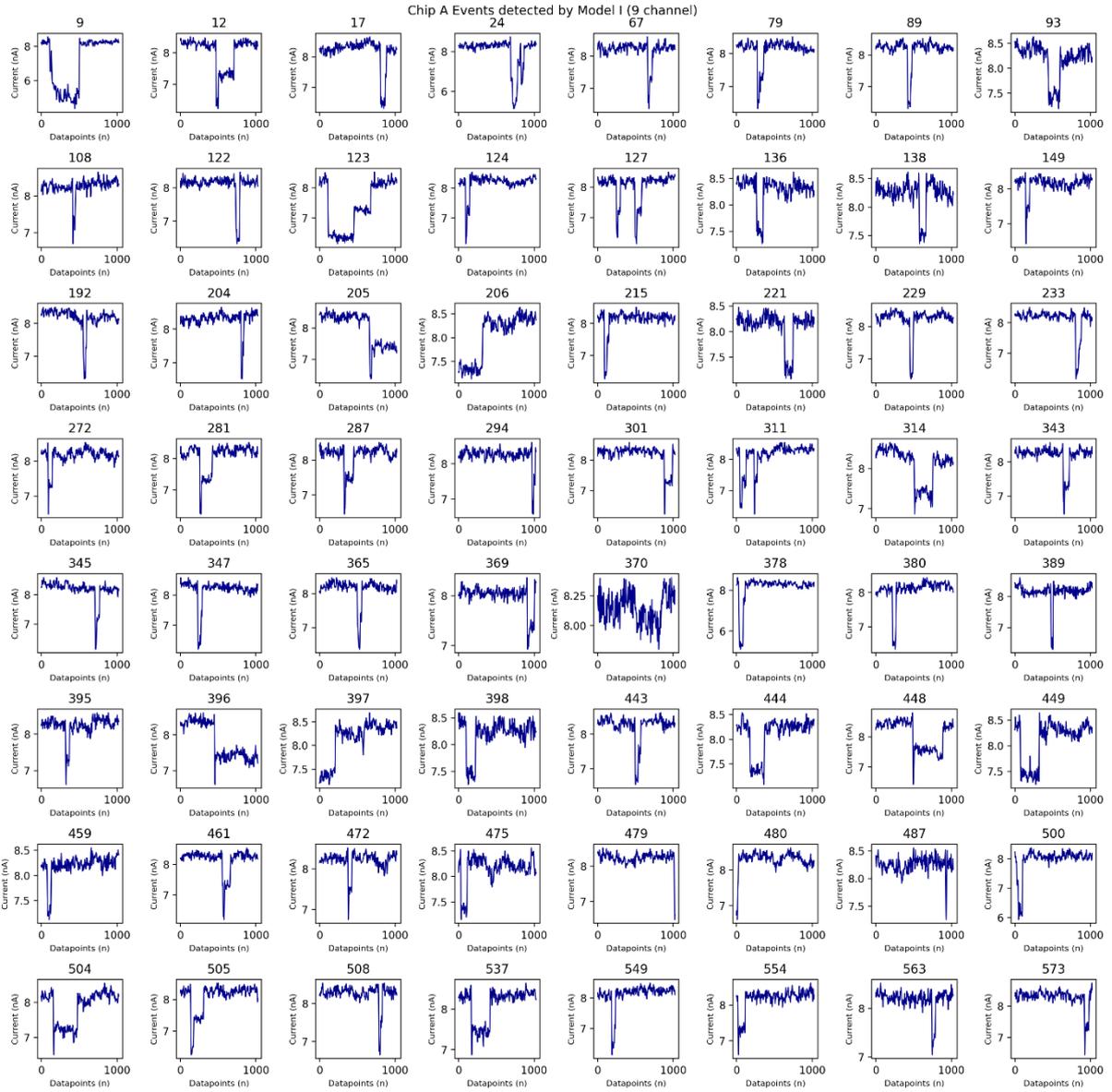

Supporting Figure SF 22: Few event bins detected by the 9-channel model from ChipA.abf file.

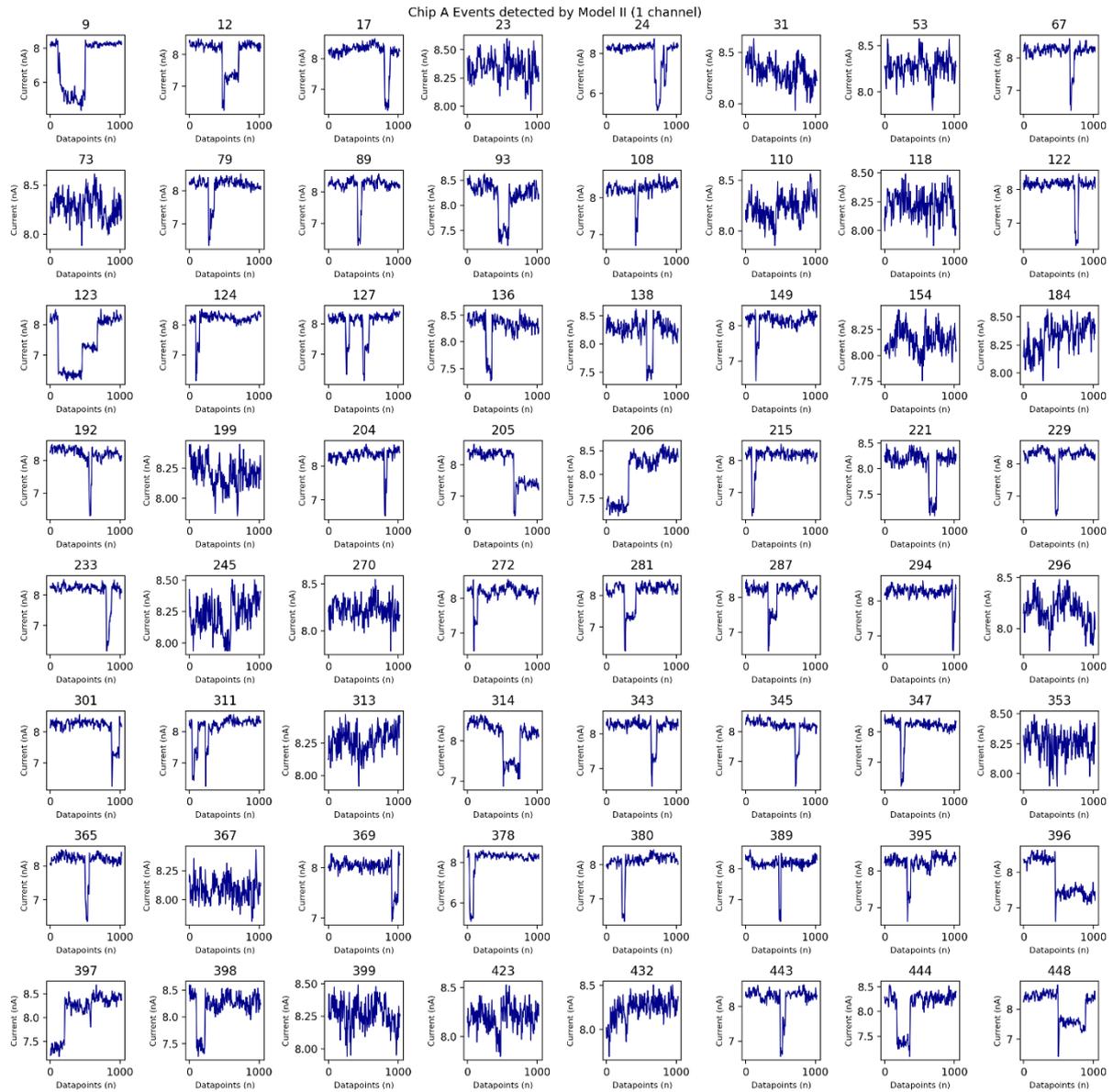

Supporting Figure SF 23: Few event bins detected by the single channel model from ChipA.abf file.

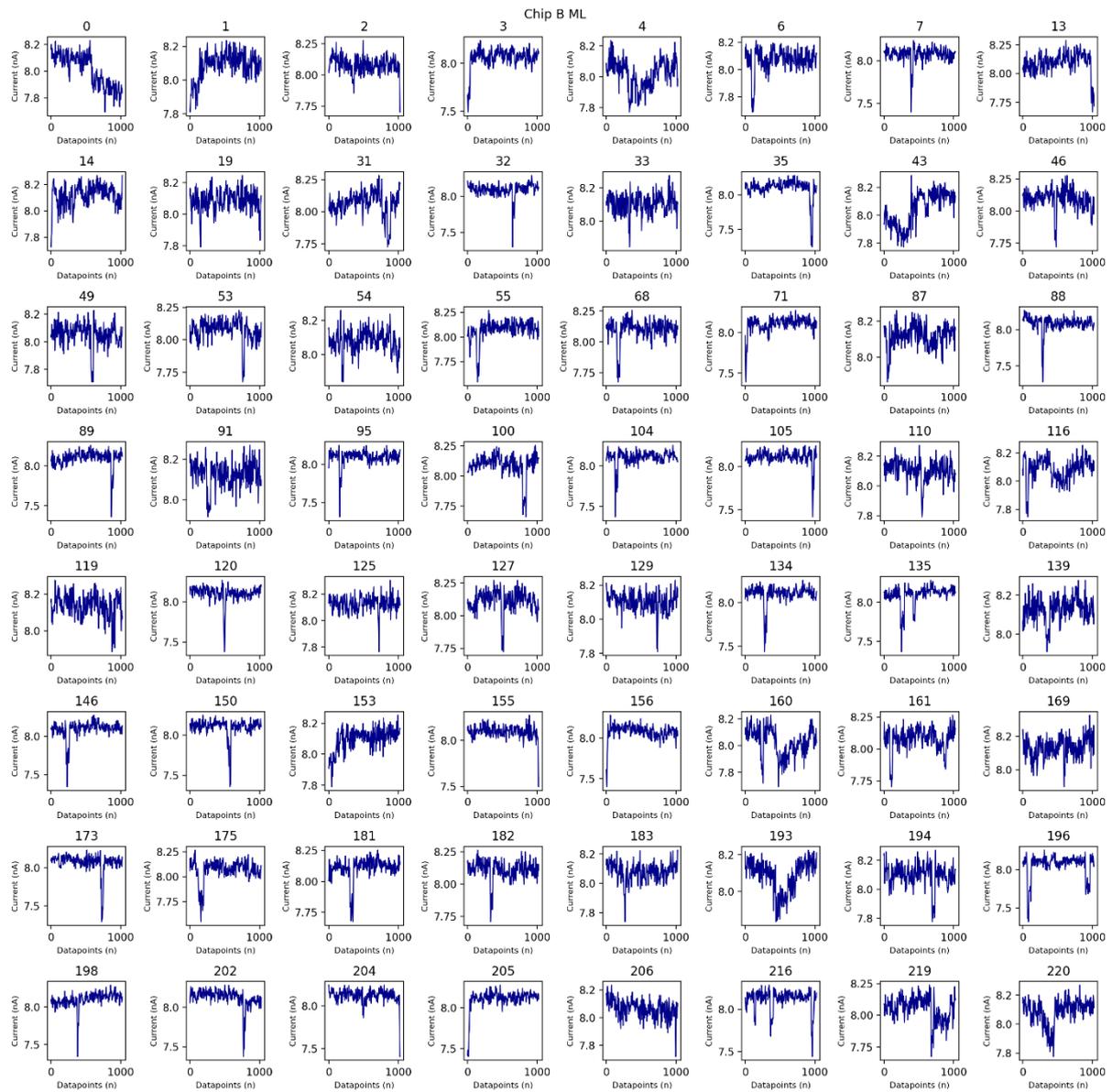

*Supporting Figure SF 24: Few event bins detected by the models (Union of events detected by single channel and 9 channel models) for the file ChipB.abf, the title of each figure represents the index of the bin.*

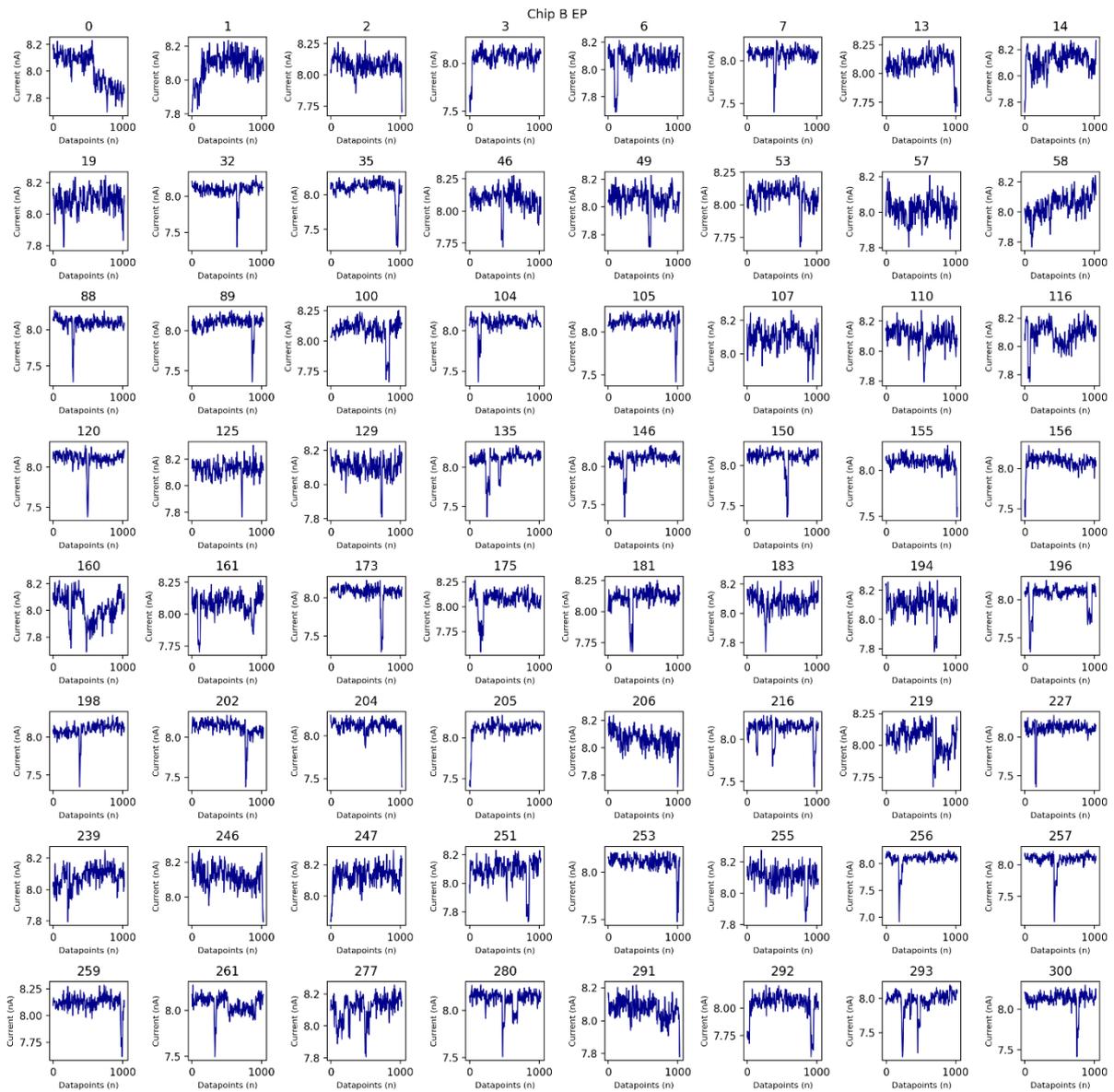

*Supporting Figure SF 25: Few event bins detected by EventPro for the file ChipB.abf, the title of each figure represents the index of the bin.*

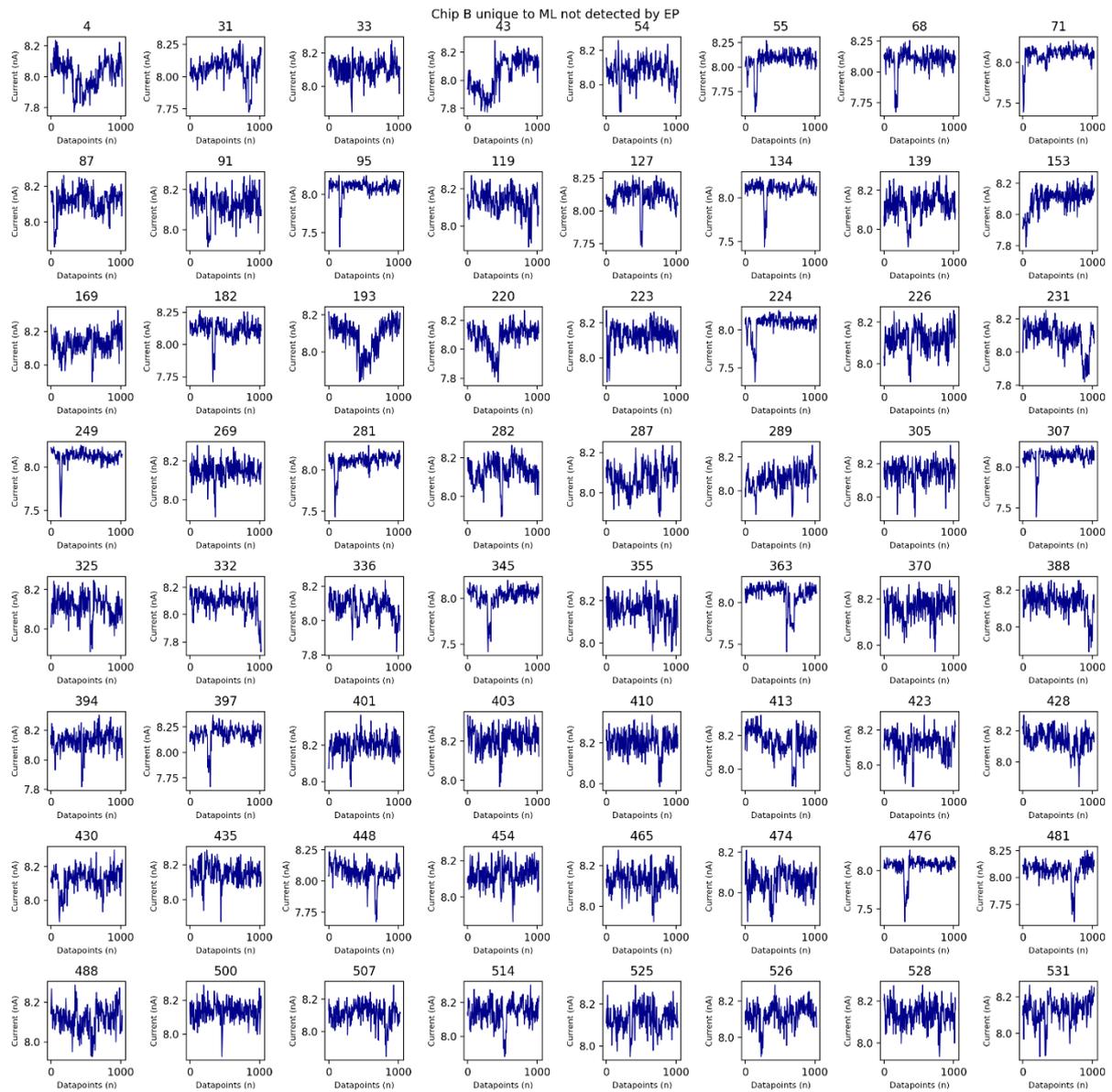

Supporting Figure SF 26: Few event bins detected by the models but not detected by EventPro for the file ChipB.abf, the title of each figure represents the index of the bin.

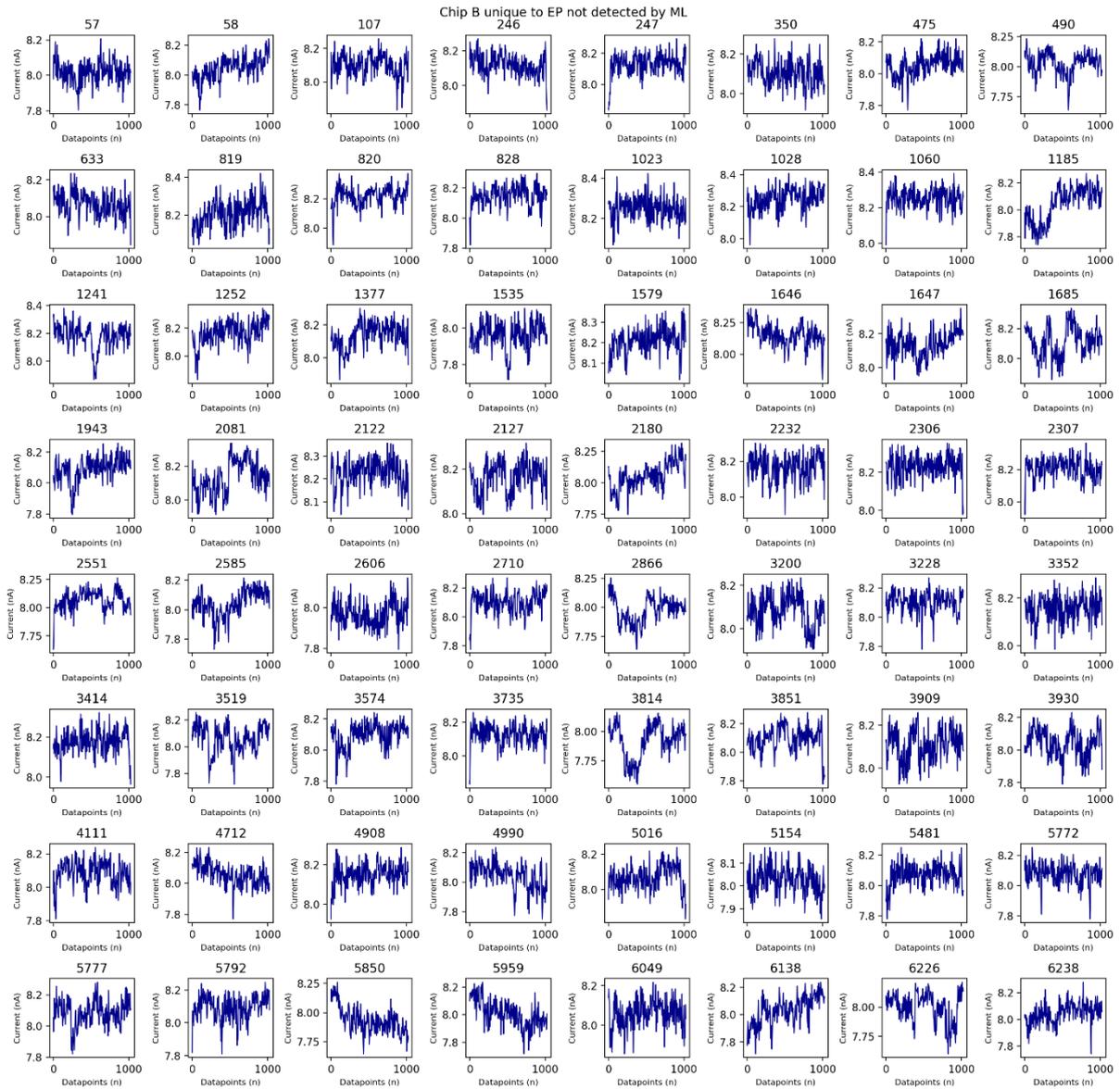

*Supporting Figure SF 27: Few event bins reported as events by EventPro, but undetected by the models for the file ChipB.abf, the title of each figure represents the index of the bin.*

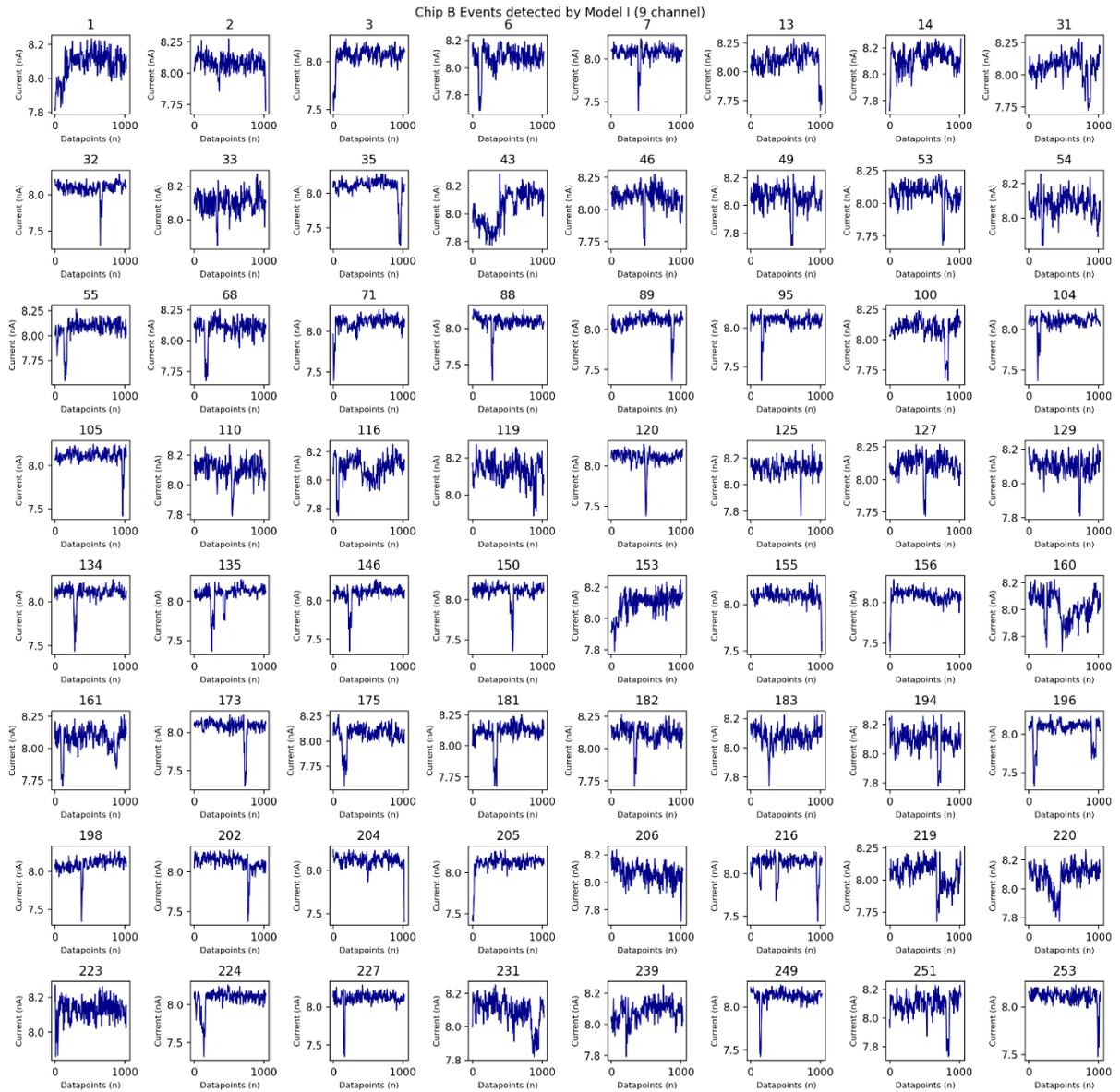

Supporting Figure SF 28: Few event bins detected by the 9- channel model from ChipB.abf file.

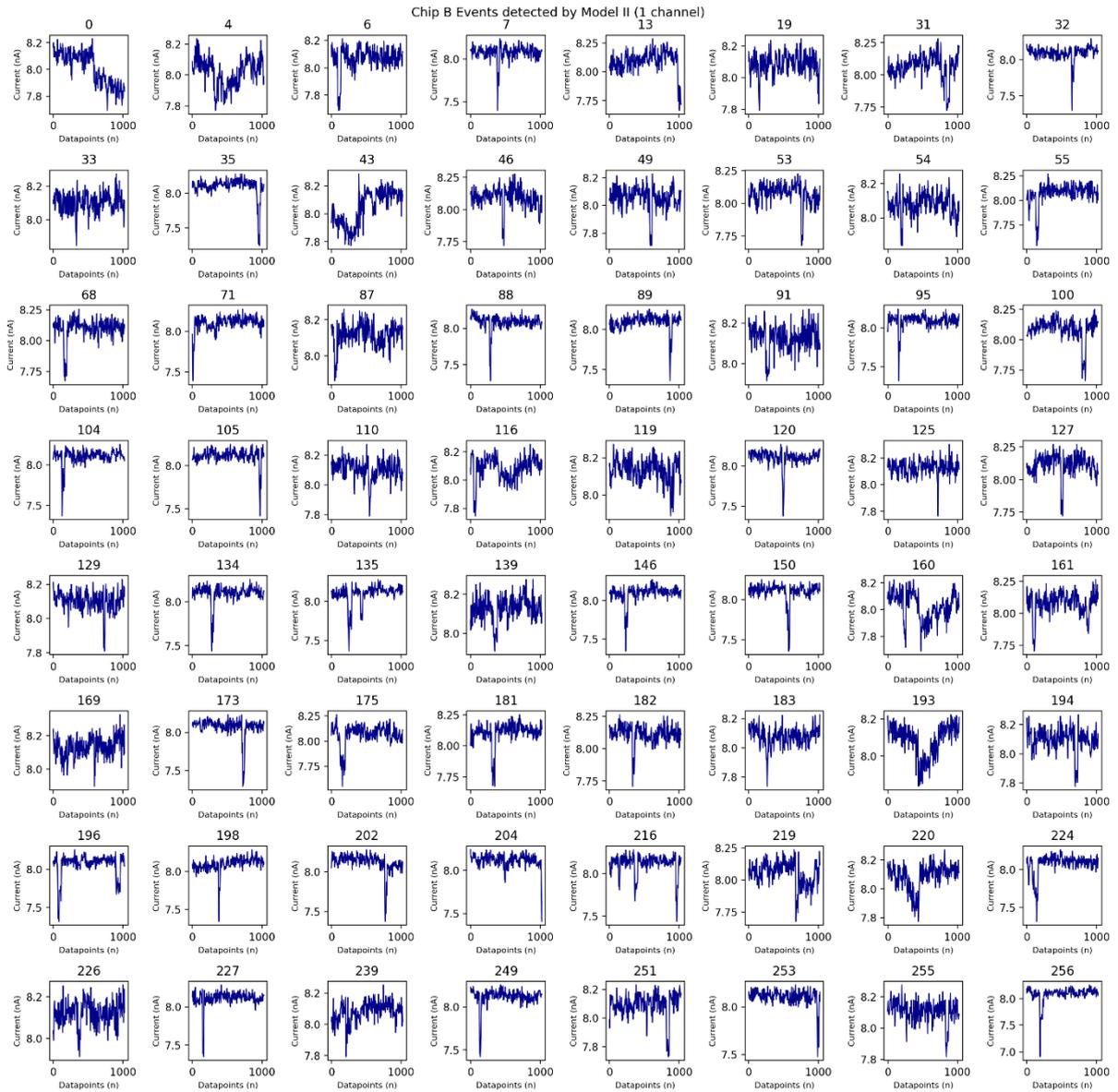

Supporting Figure SF 29: Few event bins detected by single channel model from ChipB.abf file.